\documentstyle [12pt,epsfig]{article} 
\makeatletter

\@addtoreset{equation}{section}
\makeatother

\newcommand{\ba}{\begin{eqnarray}}
\newcommand{\ea}{\end{eqnarray}}

\newcommand{\C}{{\rm C}}
\newcommand{\x}{\overline{x}}
\newcommand{\X}{\overline{X}}
\newcommand{\tb}{\overline{t}}
\newcommand{\Tb}{\overline{T}}
\newcommand{\xr}{{\rm x}}
\newcommand{\tr}{{\rm t}}
\newcommand{\Tr}{{\rm T}}
\newcommand{\Rr}{{\rm R}}

\begin{document}
\newcommand{\BS}{\bigskip}
\newcommand{\SECTION}[1]{\BS{\large\section{\bf #1}}}
\newcommand{\SUBSECTION}[1]{\BS{\large\subsection{\bf #1}}}
\newcommand{\SUBSUBSECTION}[1]{\BS{\large\subsubsection{\bf #1}}}

\begin{titlepage}
\begin{center}
\vspace*{2cm}
{\large \bf The physics of space and time III: Classification of space-time experiments
  and the twin paradox}  
\vspace*{1.5cm}
\end{center}
\begin{center}
{\bf J.H.Field }
\end{center}
\begin{center}
{ 
D\'{e}partement de Physique Nucl\'{e}aire et Corpusculaire
 Universit\'{e} de Gen\`{e}ve . 24, quai Ernest-Ansermet
 CH-1211 Gen\`{e}ve 4.
}
\newline
\newline
   E-mail: john.field@cern.ch
\end{center}
\vspace*{2cm}
\begin{abstract}
   A nomenclature for inertial frames and a notation for space and time coordinates
    is proposed to give an unambigous description of space-time experiments in
    special relativity. Of particular importance are the concepts of `base' and `travelling'
    frames and `primary' and `reciprocal' experiments. A detailed discussion of the 
     twin paradox is presented. The physical basis of the differential aging effect
     is found to be a relativistic relative-velocity transformation relation, not,
     as hitherto supposed, the spurious `length contraction' effect.
 \par \underline{PACS 03.30.+p}
\vspace*{1cm}
\end{abstract}
\end{titlepage}
 
\SECTION{\bf{Introduction}}
  The present article is the third in a series of long papers attempting to give a comprehensive
  description of the physics of flat space-time (i.e., in the absence of gravitational effects)
  by correcting certain errors in the application of the space-time Lorentz transformation (LT)
  originating in Einstein's first special relativity paper ~\cite{Ein1}. The error was the
   neglect of certain additive constants in the transformation equations, necessary to correctly
   describe synchronised clocks at different spatial positions. Although Einstein pointed out
   the necessity to include such constants in Ref.~\cite{Ein1} this was not done by Einstein
   himself or by subsequent authors. As shown by the present author in Ref~\cite{JHFLLT}, correction
   of this error renders spurious the `length contraction' (LC) and `relativity of simultaneity' (RS)
   effects of conventional special relativity theory. For further discussion of this point
    see Refs.~\cite{JHFUMC,JHFCRCS,JHFACOORD}. Although there is ample experimental verification
    of the time dilation (TD) effect, predicted using the space-time LT in Ref~\cite{Ein1}, there is,
    to date, no published experimental test of the LC or RS effects~\cite{JHFLLT}. The present
     author has proposed a test of RS using two satellites in low Earth orbit or
     GPS satellites in combination with a single satellite in low Earth orbit~\cite{JHFSEXPS}.
     \par The first of the long papers~\cite{JHFSTP1} discussed the axiomatic basis of special 
     relativity theory, recalling in an appendix two previously-published derivations of the LT
     that do not use Einstein's second postulate concerning the constancy of the speed of light.
     This paper introduces a calculus of pointer-mark coincidences with a view to giving 
     a mathematically rigorous description of experimental space and time measurements. Also
      presented are several different clock synchronisation procedures, not employing light
      signals and so not affected by considerations of `conventionality' inherent in the
     the assumption of light-speed isotropy~\cite{AAL}. 
     \par  The second long paper~\cite{JHFSTP2} is a detailed critique of Einstein's original 
        special relativity
        paper~\cite{Ein1}. As well as containing concepts and predictions that revolutionised 
       the understanding of physics, Ref.~\cite{Ein1} also contains many mistakes, both
      of a mathematical
      nature, and of physical interpretation, for example, the spurious RS and LC effects. In view
      of the importance of classical electromagnetism in the arguments presented
      in Ref.~\cite{Ein1}, the modern reassessment of this paper in Ref.~\cite{JHFSTP2} uses 
      results from work by the present author on relativistic classical
      electrodynamics~\cite{JHFPS2,JHFRSKO,JHFIND,JHFFT}
      derived as the classical limit of quantum electrodynamics.
    \par  The subject of the present paper is a study of the precise operational meaning of the coordinate 
        symbols which appear in the space-time LT, with the aim of understanding the physical
        basis of the differential aging effect (DAE) that underlies the `twin paradox' as orginally
       introduced by Langevin~\cite{Langevin}. This paradox arises from the following considerations.
       A travelling twin, T, makes a round trip to a distant star while her brother, R,
        remains on Earth. Assuming that the periods of uniform motion are much larger than those 
         of acceleration or deceleration during the journey, then, due to TD, R will see a clock
      carried by T and measuring the increment of her age, to run slow, throughout the journey,
      compared to a similar clock
      at rest in his own frame and measuring the increment of his own age during T's journey. R will
      then see that T is younger than he is on her return to Earth. However, from T's point of view,
       assuming that that the TD effect depends only on the magnitude of the relative velocity, it might be
       concluded that the clock carried
         by R will appear to T to be running slow, throughout the journey, as compared to the clock
    at rest in T's frame. Therefore, on her return, T will find that she is older than, not
        younger than, R. Since T cannot be both younger than, and older than, R, the above description
        leads to a paradox.
       \par The physical meaning of the time symbols occuring in the space-time LT is crucial
        for the resolution of the above paradox ---in fact it is the travelling twin, T, who is the
         youngest. Particularly important is the distinction, explained in Section 4 below, between 
        a primary experiment and its reciprocal as compared to a transformation of space-time coordinates and
        its inverse. Before the work presented in the present paper, its author had not appreciated
        this important distinction and its relation to the `Reciprocity Principle' (RP)~\cite{BG} which states
        that: `If the velocity of an  inertial frame S' relative to another such frame S is
        $\vec{v}$, then the velocity of S relative  to S' is  $-\vec{v}$'. As will be seen below,
        the RP, although valid in Galilean relativity, no longer holds in special relativity
        as a consequence of TD and the invariance of spatial intervals when space-time as opposed
       to kinematical (energy-momentum) LTs are considered. Earlier versions
        of several papers written 
        recently by the present author, assuming the correctness of the RP in applications of the 
        space-time LT  in special
        relativity~\cite{JHFAS,JHFTETE,JHFMUDEC,JHFSARTORI,JHFINVL} therefore contain erroneous
        arguments and must be extensively revised. To date, the general validity of the RP in special
        relativity has been universally (although usually tacitly) assumed. As will be seen
        it is the basis of the `standard solution' of the twin paradox in which the 
        physical origin of the DAE is traced to LC of spatial intervals in the frame of stay-at-home
        twin, as observed in the travelling
        twin's frame. As will be seen, the RP must be replaced by a related but distinct `Kinematical
         Reciprocity Principle' (KRP) that, unlike the RP, is valid in both Galilean and special relativity.  
       \par The structure of this paper is as follows: In the following section as an
        instructive preamble, five different ways to measure the length of a train are
        discussed. Reflecting on the assumptions underlying these methods gives the key
        to understanding the correct solution of the twin paradox. In Section 3  a nomenclature
        and notation to describe different space-time experiments is introduced. Inertial
         frames are classified according to three different criteria: subject or object, source 
         or target and base or travelling. The notation specifies each of these six attributes
         for any given coordinate. Several examples are given of applications of the notation:
         time dilation, simultaneity of events in two inertial frames,  Lorentz invariance of
        length intervals, relativistic reciprocity relations, velocity addition formulas,
        transverse and longitudinal photon clocks, Einstein's 1905 discussion of `relativity
        of simultaneity' and Einstein's train-embankment thought experiment. Section 4 
        discusses the differing concepts of inverse space-time transformations and reciprocal experiments.
         The twin paradox and its solution is discussed, in the light of the analyses
        of several different thought
         experiments, in Section 5. In Section 6 the standard solution of the  twin paradox is
         applied to the `space billiard' thought experiment introduced in Section 5 and
         contrasted with the correct solution presented in the same section. In  Section 7
        the twin paradox is discussed in relation to the Minkowski space-time plot. In this
        connection, the recent paper~\cite{JHFMinkP} by the present author pointing out an angular
        sign error in drawing the directions of the space and time coordinate axes of
       of the moving frame, originating in Minkowski's original work~\cite{Minkowski}, is 
       important. Correcting this error invalidates the standard solution of the twin
       paradox as derived from the space-time plot. Section 8 contains a brief summary
      and some closing comments.

\SECTION{\bf{Prologue: Five ways to measure the length of a train}}
 In space-time physics there is a close connection between the concepts of physical space
 and time, and that of uniform motion, that is encapsulated in Galileo's equation
 \begin{equation} 
    \Delta x = v \Delta t 
 \end{equation}
  where $\Delta x$ and $ \Delta t$ are space and time intervals respectively defined in 
   a common reference frame and $v$ is a constant velocity describing, for example, the motion
 of the origin, O', of an inertial frame, S', relative to another one, S, in the direction of a common $x$-$x'$ axis.
  Eqn(2.1) can be used to convert space measurements into time measurements and {\it vice versa }when 
   the value of $v$ is known. For example an analogue clock is based on a formula similar to (2.1)
   where the spatial interval is replaced by an angular one and the spatial velocity by a constant
   angular velocity $\omega$. 
  \par The length of a physical object (say a train) can be measured either by direct comparison
  with some other physical physical object of known spatial extent ---a ruler, or by making use
  of Eqn(2.1). The simplest way to measure the length of an object is by direct comparison of its
  ends with marks on a ruler in the rest frame of the former, the ruler also being at rest in 
  the same frame. Time plays no role in such a measurement. The spatial coincidences of the front
  and back ends of the train with the ruler marks can be observed at any times convenient for
  the experimenter. The length of the train is defined as the difference between the numbers associated
  with the ruler marks that coincide with the front and back ends of the train~\cite{JHFSTP1}. 
 \par The second method is similar, except that the measured object is in arbitary motion (i.e., in uniform,
  or accelerated, motion). The length of the train is still measured by observing the spatial coincidences 
  of the front and back of the train with ruler marks, but it is now essential that that the 
  observations are performed simultaneously in the rest frame of the ruler. Alternatively, if the
  observers at different spatial positions are equipped with synchronised clocks, and it
  is known that the motion of the train is uniform, but not the value of its speed, the length of the train
   and its speed may be determined from observations of spatial coincidences of the front and
   back of the train with ruler marks at different, but known, times~\cite{JHFSTP1}.
  \par The third method, restricted to the case of uniformly moving object with a known speed, is for a
    single observer in the rest frame of the ruler, equipped with a clock, to measure the time
     difference, $\Delta t$, between the passages of the front and back ends of the moving train past
     a fixed ruler mark. The length of the train, $\Delta x$, is then given by Eqn(2.1).
    \par The fourth method is a variant of the third
         in which the train remains at rest, say in the frame S, and the observer, equipped with
        a clock, passes by in another train, say with rest frame S', moving at a known uniform speed.
      The observer records the times at which he passes by the two ends of the train and calculates
      its length with Eqn(2.1). This method is also applicable when both the observer and the
     measured train are in motion with known speeds, when an appropriate velocity addition
    formula is used to calculate the speed, $v$, to be substituted in (2.1).
     \par The fifth method involves observation, in the frame S, of the time difference of spatial
     coincidences of the front end of the moving train, with rest frame S', used in the fourth method,
     with the back and front ends of the stationary to-be-measured train. Again the length of the train is given
    by Eq. (2.1). 
     \par Suppose now that two independent measurements of the length of a train are 
        made, one using the fourth and the other the fifth of the methods just described. 
    If different results are obtained for the length
    of the train, three different causes, in one-to-one correspondence with the quantities in Eqn(2.1),
      may be cited:
   \begin{itemize}
  \item[(i)] The train changed length between the two measurements.
   \item[(ii)] The observer's train, was moving at a different speed
              during the two measurements.
     \item[(iii)] The clocks employed in the two measurements, one in the frame S, the other
              in the frame S', are running relatively fast 
                 (or slow) with respect to each other.
    \end{itemize}
    If it now happens that one clock {\it is} running slower than the other, in the 
   comparison of the measurements, the observer, O(Slow),
   with the slower running clock, but unaware of this fact, has two possible ways to 
    interpret the different measurement results on the basis of Eqn(2.1):
 \begin{itemize}
      \item[(a)] The assumed speed is correct but the train was shorter during his
            measurement, than during that performed by the other observer, O(Fast).
       \item[(b)] The train has the same length but appears to move faster for O(Slow)
                  than for O(Fast)
       \end{itemize}
     Now if it is {\it known}, independently, both that the clock of O(Slow) does run slower
       than that of O(Fast) and that the train is the same length, only the interpretation
        (b) is possible. As will be seen in the following, this is the correct prediction
   in special relativity for the case that the clock in the frame S' runs slow, according
   to an observer at rest in the frame S, due to
   the time dilation effect. Hitherto the solution (a) relativistic `length contraction'
   has been taken as the prediction of special relativity. How this spurious result arises from
    misinterpretation of the space-time Lorentz transformation will also be explained.

 \SECTION{\bf{Nomenclature and notation for space-time experiments and some applications}} 
  The space-time Lorentz transformation relates space and time measurements recorded, predicted
  or specified in different inertial frames. In its application to any given experiment, 
  care must be taken to define the exact operational meanings of the space and time
   coordinates appearing in the transformation equations. In this connection, to avoid
 confusion, the mathematical symbols representing the coordinates must be such as to encode
  clearly the following essential information: 
\begin{itemize}
  \item[(I)] The inertial frame in which the space or time measurements are actually performed.
            i.e. the frame of the observer in the experiment.
 \item[(II)] The frame in which the primary space-time events, that undergo Lorentz transformation,
             are defined.
 \item[(III)] Whether, in the experiment in which an object is moving between 
              fixed points in an inertial frame, the observer is in the rest
              frame of such a `travelling object', or in some other inertial frame. 
  \end{itemize}
  The distinction made in (III) will be found to be of crucial importance for a correct 
  understanding of the physics underlying time dilation and the closely related
  `twin paradox', as well as the {\it modus operandi} of `photon clocks'. 
 \par In order to furnish the information mentioned in the points (I)-(III) above, the 
    following nomenclature and notation for space and time coordinates is suggested:

   \par {\it Subject frame}  \par($\vec{X}$, $T$) coordinates in observer's rest frame.
   \par   {\it Object frame} \par ($\vec{x}$, $t$) coordinates in any inertial frame in motion 
                             relative to that of the observer.
    \par{\it Source frame}  \par ($\vec{X}$, $T$); ($\vec{x}$, $t$) frame in which primary space time
                                 events are specified i.e. events subjected to Lorentz
                                  transformation.
    \par{\it Target frame}  \par ($\vec{\overline{X}}$, $\overline{T}$);
                        ($\vec{\overline{x}}$, $\overline{t}$) Lorentz-transformed coordinates.
   \par{\it Base frame}    \par($\vec{X}_B$, $T_B$); ($\vec{\overline{X}}_B$, $\overline{T}_B$) coordinates 
                                             specifying the position of an object (`travelling object')
                                              moving between fixed points in an observer's frame. 
 \par{\it  Travelling frame} \par  ($\vec{X}_T$, $T_T$); ($\vec{\overline{X}}_T$, $\overline{T}_T$)
                                        coordinates specifying outset and arrival events 
                                     (corresponding to fixed points in a base frame) in the subject frame
                                of a travelling object.

 \par It is important to note that physical predictions which change completely when base and travelling
      frames are exchanged, are, as will be seen, invariant with respect to exchange of source and target
     frames, i.e. to whether the LT or its inverse is used to perform the transformation.
     Also physical predictions do not depend on whether a frame is considered to be a 
  subject frame (capitalised symbols) or an object frame (uncapitalised symbols). This is because 
   the predictions do not depend on whether or not an observer is present to perceive them.
    The use of capitalised symbols is then optional, but may improve the clarity of
   equations. An example given below is the measurement of the lifetime of an unstable
   particle where measurements are performed uniquely in the laboratory (subject) frame
   in order to obtain a quantity defined in the rest frame (object frame) of the
   decaying particle.
 \par   In the following, some applications are given to illustrate the use of the notation.

  \par \underline{{\bf Time dilation}}
   \par Consider an experiment where clocks $\C_0$ and  $\C'_0$ are placed at the origins, O and O'
   of inertial frames S and S', where S' moves in the frame S with speed $v_B$ along the common $x$-$x'$
   axis of the two frames.  The subscript `$B$' indicates that S is the base frame for the
    travelling clock  $\C'_0$ so that $\Delta X_B(C'_0) = v_B \Delta T_B(C'_0)$. Two, distinct,
     experiments are possible: $\C'_0$  is observed from S or  $\C_0$ is observed from S'.
    If the observations are made in the frame S then the latter is the subject and base frame of the
      experiment and may be either the source or target frame for the LT. The former possibility
    is treated first, the other one will be  considered below. For this the appropriate LT,
     relating the elapsed time $\tb'(\C'_0)$ to the corresponding time, $ T(\C_0)_B$
      recorded  by $\C_0$, is:
         \begin{eqnarray}
             \x'(\C'_0)& = & \gamma_B [X(\C'_0)_B-v_B T(\C_0)_B] = 0, \\
 \tb'(\C'_0)& = & \gamma_B [T(\C_0)_B-\frac{v_B X(\C'_0)_B}{c^2}] 
        \end{eqnarray}
     where $\gamma_B \equiv 1/\sqrt{1-(v_B/c)^2}$. These transformation equations require that when O
    and O' are in spatial coincidence, $X(\C'_0)_B = 0$, then      
  $\tb'(\C'_0) = T(\C'_0)_B = 0$. Using (3.1) to eliminate 
     $X(\C'_0)_B$ from (3.2) gives immediately the time dilation (TD) relation:
     \begin{equation} 
       T(\C_0)_B = \gamma_B \tb'(\C'_0).
    \end{equation}   
      If, in the experiment with a  {\it reciprocal configuration}, $\C_0$ moves with speed -$v'_B$ relative
     to $\C'_0$  along the positive $x'$ axis, so that  $\Delta X'_B(C_0) = -v'_B \Delta T'_0(C_0)_B$
     a similar calculation gives the TD relation:
  .     \begin{equation} 
       T'(\C'_0)_B = \gamma'_B \tb(\C_0)
    \end{equation}   
     where   $\gamma'_B \equiv 1/\sqrt{1-(v'_B/c)^2}$. The TD relation for the experiment with
      a reciprocal configuration
      is obtained by exchange of primed and unprimed quantities throughout Eqn(3.3).
     \par An important feature of the TD relations (3.3) and (3.4) is their manifest translational
       invariance ---no spatial coordinates appear in these equations. The physical meaning
       of TD is that a uniformly moving clock ($\C'_0$ in (3.3)) is observed to run slower by
       the factor $1/\gamma_B$ relative to an identical clock at rest ($\C_0$ in (3.3)), and this
       independantly of the spatial position of the clock. In an identical manner the clock
       $\C_0$ is observed to run slower than $\C'_0$  by the factor  $1/\gamma'_B$ when observed from S'.
       This apparently contradictory behaviour shows that TD is a subjective (observer-dependent)
       effect. The `twin paradox', to be discussed below, arises from the attempt to relate, by the LT,
       the independent events in an experiment and the different experiment with a reciprocal
       configuration. 
   \par \underline{{\bf Simultaneity of events in two inertial frames}}
       \par If a second clock $C'_{L'}$ is at rest at $x' = L'$ in S' its equation of
      motion in the frame S is: 
   \begin{equation} 
  X(\C'_{L'})_B  = v_B T(\C_{0})+ L
 \end{equation}
   where $L$ is the separation of $\C'_{0}$ and $\C'_{L'}$ in S. 
    The space transformation equation for $\C'_{L'}$ is given by:
   \begin{equation} 
 \x'(\C'_{L'})- L'= \gamma_B( X(\C'_{L'})_B-L-v_B T(\C_{0})_B) = 0.
 \end{equation}
   This equation satisfies simultaneously the intial condition $\x'(\C'_{L'})= L'$
    and the equation of motion (3.5). It also reduces to (3.1) in the case $L = L' = 0$.
   The time transformation equation corresponding to (3.6) is
    \begin{equation} 
 \tb'(\C'_{L'}) = \gamma_B[T(\C_{0})_B - \frac{v_B(  X(\C'_{L'})_B-L)}{c^2}].
 \end{equation}
   When $T(\C_{0})_B = 0$, then (3.6) gives $X(\C'_{L'})_B = L$, independently of the value of $v_B$;
   (3.1) gives  $X(\C'_{0})_B = 0$ and (3.2) and (3.7) give $\tb'(\C'_{0}) = \tb'(\C'_{L'}) = 0$. 
     $\C'_{0}$ and $\C'_{L'}$ are therefore synchronised when $T(\C_{0})_B = 0$. 
     \par Using (3.5) to eliminate $X(\C'_{L'})_B-L$ from (3.7) gives, analogously to (3.3), the 
      TD relation:
    \begin{equation} 
   T(\C_{0})_B = \gamma_B \tb'(\C'_{L'}).
    \end{equation}
     Comparing (3.3) and (3.8) it can be seen that  $\C'_{0}$ and $\C'_{L'}$ remain 
     synchronous for all values of $T(\C_{0})_B$:
  \begin{equation} 
\tb'(\C'_{0}) = \tb'(\C'_{L'}) \equiv \tb.
  \end{equation}
   There is therefore no `relativity of simultaneity' effect for two synchronised clocks at different
    positions in the same inertial frame. That this must be so is already evident from the 
    manifest translational invariance of the TD relations (3.3) and (3.8), that have no 
      dependence on spatial coordinates. Using a simplified notation, if clock 1 in S' is compared
       with clock 2 in S, and clock 3 in S' is compared with clock 4 in S, where the spatial
       positions of all the clocks are arbitary, the corresponding TD relations are:
    \begin{eqnarray}
     T_2 & = & \gamma_B \tb'_1, \\
    \  T_4 & = & \gamma_B \tb'_3.
  \end{eqnarray}
       It follows from these equations that if $\tb'_1 = \tb'_3$ then $ T_2 = T_4$ and
    {\it vice versa} ---clocks which are synchronous in  S(S') are also synchronous in  S'(S).
  
    \par \underline{{\bf Lorentz invariance of length intervals}}

\begin{figure}[htbp]
\begin{center}\hspace*{-0.5cm}\mbox{
\epsfysize15.0cm\epsffile{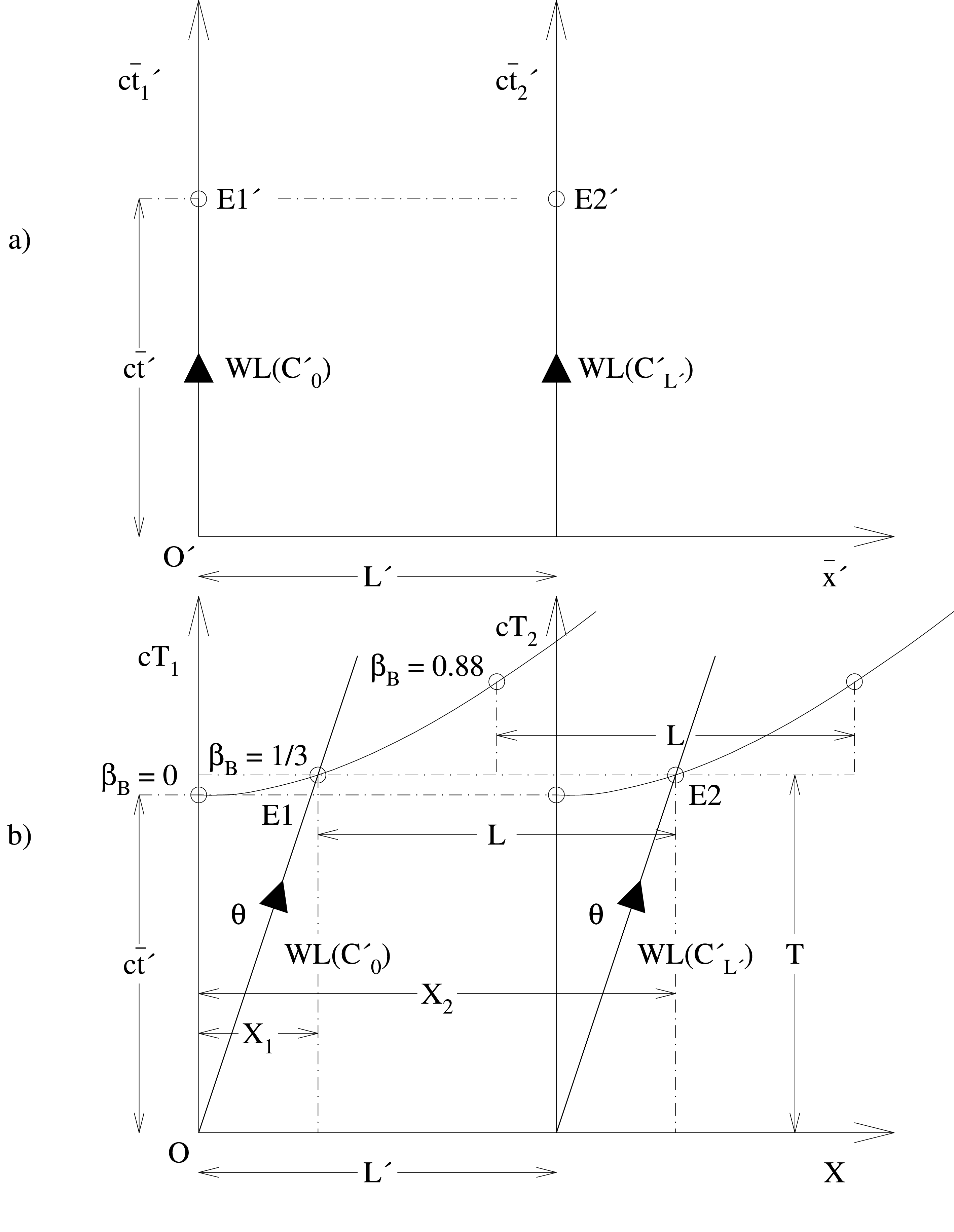}}
\caption{{\em a) World lines of the clocks, C'$_0$ and C'$_L'$, and simultaneous events,
    E1'and E2', in the frame S'. b) The same world lines and events as observed in the frame S,
    showing the invariance of the spatial separation of the clocks and the absence
   of any `relativity of simultaneity' effect. See text for discussion.}}     
\label{fig-fig1}
\end{center}
\end{figure}

     \par  Combining (3.1) with (3.3) and (3.6) with (3.8) gives the relations
    \begin{eqnarray}
      X(\C'_{0})_B & = & \gamma_B \beta_B c \tb'(\C'_{0}), \\
  X(\C'_{L'})_B & = & L + \gamma_B \beta_B c \tb'(\C'_{L'}) 
  \end{eqnarray}
    where $\beta_B \equiv v_B/c$. 
   \par In virtue of the identity $\gamma_B^2 - \beta_B^2 \gamma_B^2  \equiv 1$ (3.3),(3.12)
     and (3.8),(3.13) are two pairs of parametic equations that specify hyperbolae in the
     Cartesian space-time plot in the frame S. Introducing a simplified notation:
     $X_1 \equiv X(\C'_{0})_B$, $X_2 \equiv X(\C'_{L'})_B$, $\tb'_1 \equiv\tb'(\C'_{0})$,
      $\tb'_2 \equiv\tb'(\C'_{L'})$ and $T \equiv T(\C_{0})_B$, the equations of these 
      hyperbolae are
   \begin{eqnarray}
  c^2 T_1^2 - X_1^2 & = & c^2(\tb'_1)^2, \\
  c^2 T_2^2 - (X_2-L)^2 & = & c^2(\tb'_2)^2.
   \end{eqnarray}
     Consider now simultaneous events in S', E1' and E2', on the world lines of
       $\C'_{0}$ and $\C'_{L'}$ such that:
    \begin{equation} 
     \tb'_1 = \tb'_2 \equiv \tb',~~T_1 = T_2 \equiv T.
    \end{equation}
     The world lines of $\C'_{0}$ and $\C'_{L'}$ in S' and the events  E1' and E2'
     are shown in Fig. 1a. The transformed events in the frame S are shown in Fig. 1b, where
       the world lines of  $\C'_{0}$ and $\C'_{L'}$ for $\beta_B = 1/3$ are drawn. Also shown are
       are the hyperbolae (3.14) and (3.15) corresponding to the simultaneity condition (3.16)
       together with their intersections with the world lines of $\C'_{0}$ and $\C'_{L'}$
       for $\beta_B = 0$, $\beta_B = 1/3$ and  $\beta_B = 0.88$. For $\beta_B = 0$ the world lines are identical
        to those in Fig. 1a, lying along the $c\tb'_1$ and $c\tb'_2$ axes. Noting that with 
        simultaneity condition (3.16), $X_1$ and $X_2$ are functions only of $\beta_B$, 
        (3.14)-(3.16) give the relation:
 \begin{equation}
         X_2(\beta_B) - X_1(\beta_B) = L
    \end{equation}
   where $L = X_2(T = 0)$ is independent of $\beta_B$. Eqn (3.17) holds for all values of 
    $\beta_B$, in particular for $\beta_B \rightarrow 0$ when  $X_1 \rightarrow X'_1$
     and $X_2 \rightarrow X'_2$ so that
   \begin{equation}
       X_2(0) - X_1(0) = X'_2 - X'_1 \equiv L' = L.
     \end{equation}
    As is also clear from inspection of Fig. 1b, the spatial separation of 
 $\C'_{0}$ and $\C'_{L'}$ is a Lorentz-invariant quantity ---there is no `relativistic length contraction'.
     How this spurious effect arises from misuse of the space-time LT is explained in Section 6 below.

    \par \underline{{\bf Relativistic reciprocity relations}}
    \par The TD effect and the invariance of length intervals imply modification
     of the velocity reciprocity relation of Galilean relativity, which may be stated as:
    \begin{equation}
      \left. \frac{d X_B}{dT_B}\right|_{\X'_T} \equiv v_B =
      - \left.  \frac{d \X'_T}{d \Tb'_T}\right|_{X_B}  \equiv \bar{v}'_T.~~~(\rm{Galilean~Relativity})  
    \end{equation}
       The bar in the symbol $\bar{v}'_T$ denotes that, unlike $v_B$, which is a fixed input parameter,
       the quantity is a derived one analogous the the transformed coordinates $\bar{x}$, $\bar{t}$.
       A similar notation is used below  to classify other kinematical parameters.
      In Galilean relativity
      $\Delta X({\rm O}')_B = -\Delta \X'({\rm O})_T$, $\Delta T_B = \Delta \tb' = \Delta \Tb'_T$.
       For an observer at rest in S'
     (a `travelling' observer if S is the base frame), $\Delta \tb'$ becomes a 
      subject frame interval. Because of the invariance of length intervals, the relation
    $\Delta X({\rm O}')_B = -\Delta \X'({\rm O})_T$ holds also in special relativity,
     whereas, due to TD, time intervals
   in S and S' are related in a different manner than in Galilean relativity:
      $\Delta T_B = \gamma_B \Delta \tb' = \gamma_B \Delta \Tb'_T$. Combining these relations
      between infinitesimal intervals, the reciprocity relation 
       in special relativity therefore differs from (3.19):
      \begin{equation}
      \left. \frac{d X_B}{dT_B}\right|_{\X'_T} \equiv v_B =  - \left.  \frac{d \X'_T} {\gamma_B d \Tb'_T}\right|_{X_B}  \equiv
     \frac{\bar{v}'_T}{\gamma_B}.~~~(\rm{Special~Relativity}) 
    \end{equation}
    So that 
       \begin{equation}
        \bar{v}'_T = \gamma_B v_B.
       \end{equation}
    As will be seen below, the relation (3.21) is crucial for understanding the correct solution 
   of the twin paradox.
    \par Consideration of the experiment, with a reciprocal configuration, in which S' is the base frame
     and $\C_0$ is 
    the travelling object yields the following reciprocity relation:
        \begin{equation}
      \left. \frac{d X'_B}{dT'_B}\right|_{\X_T} \equiv- v'_B = - \left.  \frac{d \X_T} {\gamma'_B d \Tb_T}\right|_{X'_B}  \equiv
     -\frac{\bar{v}_T}{\gamma'_B}
 \end{equation}
     giving
      \begin{equation}
        \bar{v}_T = \gamma'_B v'_B.
       \end{equation}
    The relation (3.22) is obtained from (3.20) by exchange of primed and unprimed quantities
    followed by the operations $ v'_B  \rightarrow - v'_B $, $ \bar{v}_T  \rightarrow - \bar{v}_T $. Eqn(3.23)
      is related to (3.21) simply by exchange of primed and unprimed
    quantities. 

   \par \underline{{\bf Velocity addition formulas}}
 
   \par  The addition of base frame velocities is first considered. Suppose that the frame S'
      moves with speed $v_B$ in the positive $x$-direction and that an object, O, moves with 
       speed $u'_B$ in the frame S' in the same direction. Having fixed the values of $v_B$
        and $u'_B$, what is the speed, $\bar{w}_B$, of O in the frame S? The differential form
       of the LT between S and S' is:
       \begin{eqnarray}
        d\bar{x}'({\rm O})_B & = & \gamma_B[ dX({\rm O})_B -v_B dT({\rm O})_B],  \\
   d\bar{t}'({\rm O})_B & = & \gamma_B[ dT({\rm O})_B -\frac{v_B dX({\rm O})_B}{c^2}].
      \end{eqnarray}
    Since both S and S'are base frames for the travelling object O,
    \begin{eqnarray}
           \frac{d\bar{x}'({\rm O})_B}{ d\bar{t}'({\rm O})_B} & \equiv & u'_B, \\
        \frac{d X_B({\rm O})_B}{ d T_B({\rm O})_B} & \equiv & \bar{w}_B
 \end{eqnarray}
   the base frame velocity addition formula given by combining (3.24)-(3.27)
     is the conventional parallel velocity addition formula of special relativity
    as derived by Einstein~\cite{Ein1}:
   \begin{equation}
        \bar{w}_B = \frac{u'_B + v_B}{1+\frac{u'_B v_B }{c^2}}.
   \end{equation}      
      Notice that in this derivation no travelling frame coordinates appear, only
      base frame or transformed base frame coordinates of the travelling object
      O in the frames S or S'. 
       \par Consider now a different experiment in which the object O is initially
        defined to move with speed $w_B$ along the positive $x$-axis in S, and is subsequently
        observed in the travelling frame S'. If the origin of S, the origin of S' and O coincide at time $T_B = 0$,
          then at the later time,  $T_B$, the separation, in S, of O from the origin of S' is:
  \begin{equation}
 \Delta X({\rm O})_B = (w_B-v_B)T_B.
   \end{equation} 
  The corresponding travelling frame (S') separation (c.f. Eqn (3.20)) is:
 \begin{equation}
 \Delta \bar{X}'({\rm O})_T = \bar{u}'_T  \bar{T}'_T
  \end{equation} 
 where $\bar{u}'_T $ is the speed of O in S'. 
    The Lorentz invariance of length intervals $\Delta X({\rm O})_B = \Delta \bar{X}'({\rm O})_T$
  and the TD relation 
       $T_B = \gamma_B  \bar{T}'_T$ then give the velocity transformation formula:
   \begin{equation}
      \bar{u}'_T = \gamma_B(w_B-v_B).
  \end{equation}
   When $w_B = 0$, $ \bar{u}'_T \rightarrow -\bar{v}'_T$ 
     and the relation (3.21) is recovered.
  \par It is interesting, in view of the discussion of `photon clocks' in the following
     section, to also consider the velocity transformations for an object moving in the x-y plane.
     Introducing the $x$- and $y$-components of $u_B$ and $\bar{w}_B$, the following
      formulas are obtained from (3.24), (3.25) and the invariance of tranverse coordinates
      under the LT:
      \begin{eqnarray}
       \bar{w}_B^{(x)} & = & \frac{(u_B^{(x)})'+v_B}{1+\frac{(u_B^{(x)})' v_B }{c^2}}, \\ 
       \bar{w}_B^{(y)} & = & \frac{(u_B^{(y)})'}{\gamma_B(1+\frac{v_B (u_B^{(x)})'}{c^2})}.
       \end{eqnarray}
     These are the usual relativistic formulas for the transformation of these
     velocity components.
     When $(u_B^{(x)})' = -v_B$ then (3.32) gives $ \bar{w}_B^{(x)} = 0$ and (3.33) gives:
   \begin{equation} 
 \bar{w}_B = \bar{w}_B^{(y)} =  \frac{(u_B^{(y)})'}{\gamma_B(1-\frac{v_B^2}{c^2})}
 = \gamma_B (u_B^{(y)})'
 \end{equation}
   which is similar to the transformation law (3.23) of longitudinal
    relative velocity between a base frame S' and a travelling frame S on making the
    replacements: $\gamma'_B \rightarrow \gamma_B$, $v'_B \rightarrow (u_B^{(y)})'$ and $\bar{v}_T   
   \rightarrow \bar{w}_B^{(y)}$ . Notice that in (3.34)
    $\bar{w}_B^{(y)}$ is  now the total velocity of O relative to the frame S.
   For purely transverse motion in the frame S there is therefore no distinction between
   transformation formulas for base frame and relative velocities.

   \par \underline{{\bf Transverse and longitudinal photon clocks}}
     \par In the discussion of TD above, the frame S was subject, source and base frame. In the case
    of `photon clocks' the primary space time events in the problem are defined by spatial coincidences
     between light signals and stationary objects in the rest frame of the `clock', which are then
     observed from a frame in uniform motion relative to the clock frame. Events on the world
      lines of the clocks $\C'_0$ and $\C'_{L'}$ introduced above, corresponding to light signal
      coincidences, are then the primary events in the source and travelling frame S', while S 
      remains the subject and
      base frame of the experiment, but is now the target frame of the LT. The appropriate LT
      for the problem is then the inverse of Eqns(3.1) and (3.2) with the replacements:
      $X,T \rightarrow \X,\Tb$, $\x,\tb \rightarrow x, t$:
        \begin{eqnarray}
             \X(\C'_0)_B& = & \gamma_B [x'(\C'_0)+v_B t'(\C_0)],  \\
        \Tb(\C_0)_B& = & \gamma_B [t'(\C'_0)+\frac{v_B x'(\C'_0)}{c^2}]. 
        \end{eqnarray}
       Since $x'(\C'_0) = 0$, the TD relation is given directly by Eqn(3.36):
      \begin{equation}
  \Tb(\C_0)_B  =  \gamma_B t'(\C'_0)
       \end{equation}
       which is related to (3.3) above simply by exchange of source and target frames.
       \par The use of (3.37) to analyse transverse and longitudinal `photon clocks',
       without further explicit use of the LT, is illustrated in Fig. 2 and Fig. 3 respectively.
        In order to render the photon coincidence events, defined in the frame S', visible
        to an observer in S, the photon sources and mirrors are equipped with light detectors
        (not shown) and lamps that flash when light signals are emitted, reflected or return
        to their source. 
        In the figures and related text an evident abbreviated notation is used.
        For example $\Tb(C_0)_B \rightarrow \Tb$, $t'(\C'_0), t'_{L'}(\C'_{L'})\rightarrow t'$,
        denoting the times recorded by a synchronised clock at any position in the frames S and S'
        respectively, and $v_B \rightarrow v$,$\gamma_B \rightarrow \gamma$. In Figs. 2 and 3,
        $v = (\sqrt{3}/2)c$, $\gamma = 2$. 
        \par In Fig. 2a a directed light signal (for example a laser beam) is emitted at $t0' = 0$ by
         the source So at rest in the frame S',
         and the associated lamp L$_{{\rm So}}$ flashes. The signal from L$_{{\rm So}}$ is observed 
        in the frame S
         at time $\Tb0 = 0$ (Fig. 2d). In the frame S, the directed light signal propagates at an
         angle $\theta = {\rm arcos}[v/c]$ relative to the positive $x$-axis. At time $t1' = {\rm L}/c$
           in S' the light signal arrives at the mirror, M,  where it is detected and partially reflected
          back towards the source (Fig. 2b); the lamp L$_M$ fires after the reflection
          event with a negligibly small delay.
          According to Eqn(3.37) the signal from L$_{{\rm M}}$ is observed in S at time
           $\Tb1 = \gamma  L /c$ when the source is at $\X =  v\gamma  L /c$ (Fig. 2e).
           As shown in Fig. 2c, the reflected part of the light signal arrives back at the
           source at time  $t2' = 2 L/c$  and the lamp  L$_{{\rm So}}$ flashes a second time.
           This flash is observed in S at time  $\Tb2 = 2 \gamma  L/c$ when the source is
            at $\X = 2v\gamma L/c$ (Fig. 2f). 
            The apparent speed of the directed light signal in S as measured from
            the time interval between observation of L$_{\rm M}$ and either the first or second observation
            of L$_{{\rm So}}$, is:
            \begin{equation}
             \bar{v}_{\bot}^{app} = \frac{cD}{\gamma {\rm L}} = \frac{c}{\gamma{\rm L}}\left(\frac{v \gamma {\rm L}}
                {c \cos \theta }\right) = c
             \end{equation}
              where (see Fig. 2) $D =$ PQ $=$ PR.
        Note that (3.38) is consistent with the transformation formula (3.33). Setting $(u_B^{(x)})' = 0$,
        $(u_B^{(y)})' = c$ gives  
   \begin{equation}
       \bar{w}_B^{(y)} = \frac{c}{\gamma}
  \end{equation}
       so that
  \begin{equation}
  \Tb2 = \frac{2 L}{\bar{w}_B^{(y)}} =\frac{2\gamma L}{c} 
 \end{equation}
   in agreement with the above calculation.
           \par The corresponding sequence of events for a longitudinal photon clock is shown in Fig. 3.
            Figs. 3a, 3c and 3e show the emission, reflection and return-to-source events in S', 
            while Figs. 3b, 3d and 3f show the same sequence of events in S, as observed via the
           prompt signals of the lamps  L$_{So}$ and L$_M$, in a similar manner to that described
           above for the transverse photon clock. Note that, due to the relation (3.18), and contrary
           to text-book special relativity, the separation of So and M is the same in the frames S 
           and S'. The geometry of  Fig. 3d shows that the apparent speed of the light signal
           in S', as viewed from S, obtained by observation of the time difference between the
         first signal from L$_{So}$ and
           the signal from L$_M$ is:
  \begin{equation}
          \bar{v}_+^{app} = \frac{L(1+\beta \gamma)}{\gamma L/c}
              = c \left(\frac{1}{\gamma}+ \beta\right)
  \end{equation} 
        where $\beta \equiv v/c$, Similarly the apparent light speed for the return path
         given by time difference between the observations in S of the signal from L$_{{\rm M}}$
     and the second signal
         from  L$_{{\rm So}}$ is given by the geometry of Fig. 3f as:
 \begin{equation}
          \bar{v}_-^{app} = \frac{L(1-\beta \gamma)}{\gamma L/c}
              = c\left(\frac{1}{\gamma}- \beta\right).
  \end{equation} 
     These formulas are consistent with the longitudinal relative velocity transformation since
    formula (3.41) can be written as:
 \begin{equation} 
     c = u'_+  = \gamma(\bar{v}_+^{app} -v)
  \end{equation} 
   and (3.42) as
 \begin{equation} 
     c = u'_-  = \gamma(\bar{v}_-^{app} +v).
  \end{equation}    
  The relative velocity as, observed in S, of the photon and the moving photon clock
  is $\bar{v}_+^{app} -v$ on the outward transit and $\bar{v}_-^{app}+v$ on the return
  one.

 \begin{figure}[htbp]
\begin{center}\hspace*{-0.5cm}\mbox{
\epsfysize12.0cm\epsffile{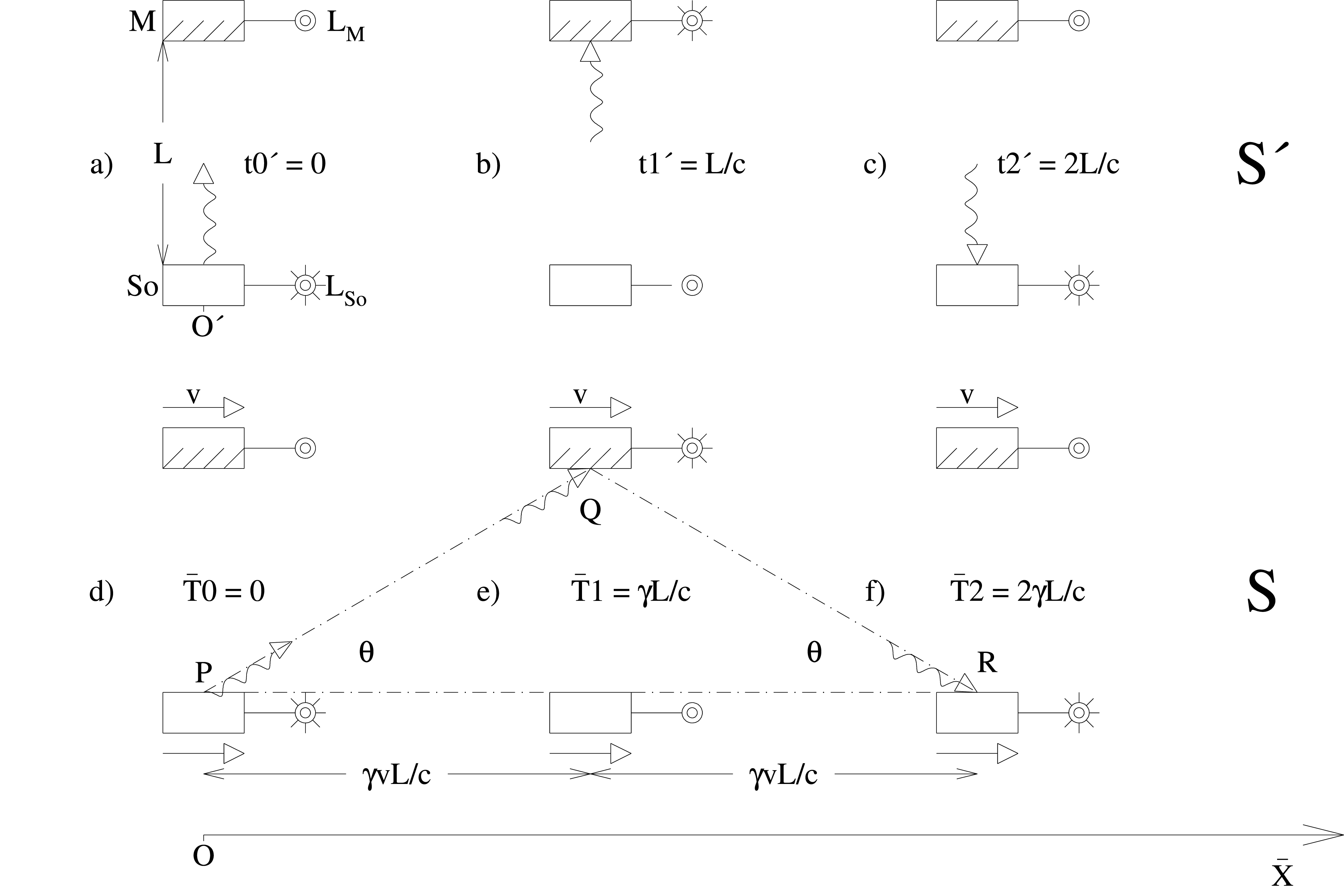}}
\caption{{\em A transverse `photon clock'. a) a directed light signal
    is emitted from the source So at rest in S' towards the mirror M;
    the lamp L$_{{ \rm So}}$ flashes. b) the light signal arrives at M;
    the lamp L$_{{ \rm M}}$ flashes. c) the reflected signal arrives
    back at  So and L$_{{\rm So}}$ flashes a second time. d),e) and f)
    show the same sequence of events as observed in the frame S,
    where S' moves with speed $v$ along the positive $\bar{X}$ axis.
     $v =(\sqrt{3}/2)c$, $\gamma = 2$.}}     
\label{fig-fig2}
\end{center}
\end{figure}

\begin{figure}[htbp]
\begin{center}\hspace*{-0.5cm}\mbox{
\epsfysize15.0cm\epsffile{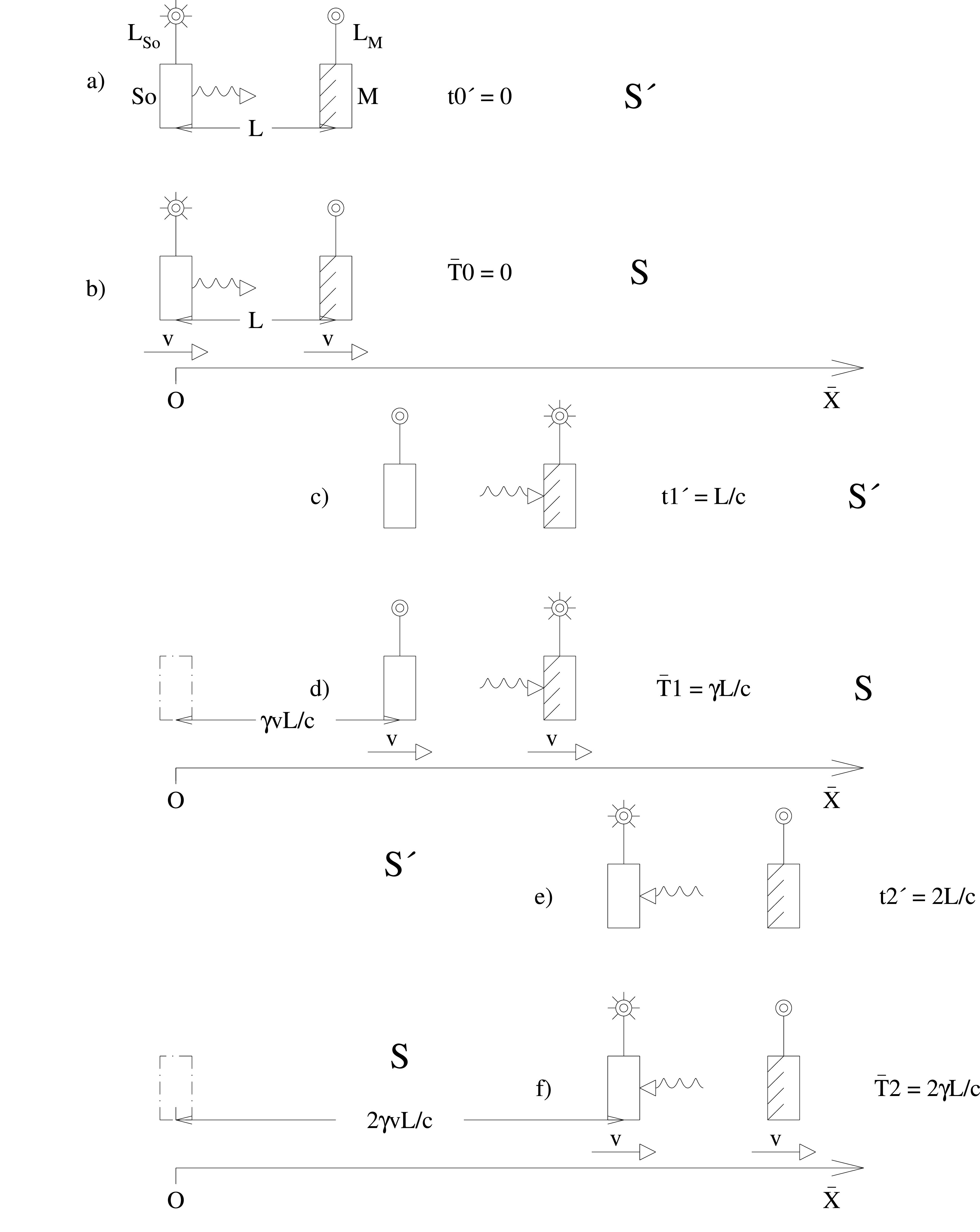}}
\caption{{\em A longitudinal `photon clock'. The event sequence in S'
    is similar to that in Fig. 1. a), c) and e) show event
    configurations in the frame S'.  b), d) and f) the same events as
    observed in S. $v =(\sqrt{3}/2)c$, $\gamma = 2$. Note that, with this choice
    of $v$, $\bar{v}_-^{app}$ is negative; i.e. the apparent velocity
     of the photon on the return path is directed in the positive $\X$ direction.}}
\label{fig-fig3}
\end{center}
\end{figure}

    \par In text books and the pedagogical literature, following the premises of Einstein's
     1905 paper~\cite{Ein1} it is universally, but incorrectly, assumed that the apparent speed
      of the light signals in the longitudinal photon clock, as viewed from S, is $c$, as is 
      indeed the case, as Eqn(3.38) shows, for the transverse photon clock. Also, events defined
      uniquely in the frame S are considered. In fact S is both the subject and source frame, but
      no object frame coordinates appear in the calculation since no LT is performed. These assumptions
      lead to the following S-frame times for the reflection and return-to-source events:
       \begin{eqnarray}
         T_{refl}& = & \frac{L}{c-v}, \\
 T_{retn}& = & \left[\frac{L}{c-v}+ \frac{L}{c+v}\right].
       \end{eqnarray}
      Note that, unlike for the case of the correct application of the LT in Fig. 3,
     $T_{retn} \ne 2 T_{refl}$. The translational invariance of the TD relation 
       then shows that $ T_{refl}$ and $T_{retn}$ cannot be related by the LT with the
       corresponding events defined in the frame S' ---equal time intervals between events in the 
        frame S' must correspond to equal time intervals between the same events viewed in the frame
           S. This is a necessary consequence of the TD relations (3.3) and (3.5).
        Making the further assumption, in contradiction with the invariant relation, (3.18), that
  the spatial separation of So and M is different in S and S' and applying the TD relation
    (3.37) to the return-to-source event gives:
    \begin{equation}
    T_{retn}= \frac{2 \gamma L'}{c} = \left[\frac{L}{c-v}+ \frac{L}{c+v}\right]
              =  \frac{2 L c}{c^2-v^2}  
  \end{equation} 
      from which follows the spurious `length contraction' relation:
    \begin{equation}
       L =  L'/\gamma.
  \end{equation} 
     The crucial error in this calculation is that (3.47) has been obtained, not by Lorentz 
      transformation of the light signal-source coincidence event in the frame S' into the frame S,
      but by considering only events in the frame S obtained on the incorrect assumptions that: Firstly,
       the apparent speed of light signal defined is S' is $c$ in the frame S, and secondly, that these
        events {\it are} related to the corresponding events in S' by the LT (the first member
         of Eqn(3.47)).  Correct application of the LT shows, viz Eqn(3.41) and (3.42),
       that the both of these assumptions are untrue. As will now be discussed, Similar mistakes 
      occur in the discussion of `relativity of simultaneity' in Einstein's 1905
       special relativity paper~\cite{Ein1}, as well as in Einstein's interpretation of
     the train embankment thought experiment~\cite{EinTETE}.

 \par \underline{{\bf Einstein's 1905 discussion of `relativity of simultaneity'}}
 \par In the original special relativity paper~\cite{Ein1}, a `photon clock' was
  used as the basis for Einstein's light signal clock synchronisation procedure' (LSCSP).
  Refering to the longitudinal photon clock shown in Fig. 3, and associating clocks
  $\C'_0$ and $\C'_L$ with So and M respectively, the LSCSP states that that these clocks
  are synchronous in the frame S' provided that:
   \begin{eqnarray}
      t1' -t0'& = & t2'-t1', \\
      \frac{2L}{ t2'-t0'} & = & c.
 \end{eqnarray}
   Einstein then considered the times of spatial coincidences of light
   signals specified in the frame S with the moving clocks  $\C'_0$ and $\C'_L$ in this frame,
   as observed in S, as discussed above in connection with Eqns(3.45) and (3.46). S is thus
   both subject and source frame ---no space time transformations are considered.
   Introducing a similar notation for sucessive coincidence events as in Figs. 2 and 3:
   $\Tb0 = 0$, $T_{refl} = T1$ and $T_{retn} = T2$ equations similar to those of \S 2
   of Ref.~\cite{Ein1} are obtained:
       \begin{eqnarray}
        T1 -T0 & = & \frac{L}{c-v}, \\
        T2 -T1& = & \frac{L}{c+v}.
       \end{eqnarray}
    Since $T1-T0 \ne T2-T1$, Einstein concluded that the clocks that are synchronous according to
    the LSCSP in S', are not so in S because the times in this frame do not satisfy the condition
     (3.49). The tacit assumption made here (though at this stage of the presentation
      in Ref.~\cite{Ein1} the LT has not yet been introduced) is that $T0$, $T1$ and $T2$ are
     related to $t1'$, $t2'$ and $t3'$
      respectively by the space-time transformation equations of special relativity. The explicit
      results given above show that this is not the case. In fact the correctly transformed times
      $\Tb0$, $\Tb1$ and $\Tb2$ {\it do} satisfy the  LSCSP condition (3.49):
      \begin{equation}
      \Tb1 -\Tb0 = \Tb2 - \Tb1
      \end{equation}
      --- the clocks are indeed also synchronous in S according the LSCSP. However the times in S
       respect not the condition (3.50) but instead
    \begin{equation}
      \frac{2L}{T2-T0}  = \frac{c}{\gamma}.
      \end{equation}
       The average speed of the light signals in S', as viewed from S, is reduced by the factor
         $1/\gamma$, consistent with a universal TD effect: all observed physical processes in
        the travelling frame S' ---including
        light propagation--- being slowed down in the same manner.
        \par Einstein's conclusion is therefore based on two false assumptions:
        \begin{itemize}
       \item[(1)] The times $T1$ and $T2$ are the Lorentz transforms of $t1'$ and $t2'$.
       \item[(2)] The apparent speed in S of light signals defined in S' is $c$.
        \end{itemize} 
      In fact, the times $T1$ and $T2$ are specified by spatial coincidences of light signals
       with M and So uniquely in the frame S. They are not connected by the LT with the similarly
       defined coincidence times  $t1'$ and $t2'$ in the frame S'. As shown by Eqs.(3.38),(3.41)
       and (3.42),
       the assumption (2) is correct for a transverse photon clock but not for
       a longitudinal one relevant to the present problem. 
      
      \par \underline{{\bf Einstein's train-embankment thought experiment}}

   \par In the popular book `Relativity, the Special and General Theory' Einstein introduced
    a thought experiment~\cite{EinTETE} intended to illustrate, in a simple way, `relativity of
     simultaneity', without consideration of the LT,
     that is frequently discussed in text books and the pedagogical literature~\cite{TETE1,TETE2,TETE3}.
      Light signals are produced by lightning strikes which simultaneously hit an embankment 
     at positions coincident with the front and back ends of a moving train. The signals are
     seen by an observer, O$_T$, at the middle of the train and an observer,  O$_B$, on the
     embankment, aligned with
     O$_T$ at the instant of the lightning strikes. The light signals are observed simultaneously
     by  O$_B$ who concludes that the lightning strikes are simultaneous. Because of the
     relative motion of O$_T$ and the light signals, the latter are not observed by O$_T$
     at the same time. Invoking the constancy of the speed of light in the train frame, Einstein
     concludes that O$_T$ would not judge the strikes to be simultaneous, giving rise to a
     `relativity of simultaneity' effect between the train and embankment frames.

\begin{figure}[htbp]
\begin{center}\hspace*{-0.5cm}\mbox{
\epsfysize10.0cm\epsffile{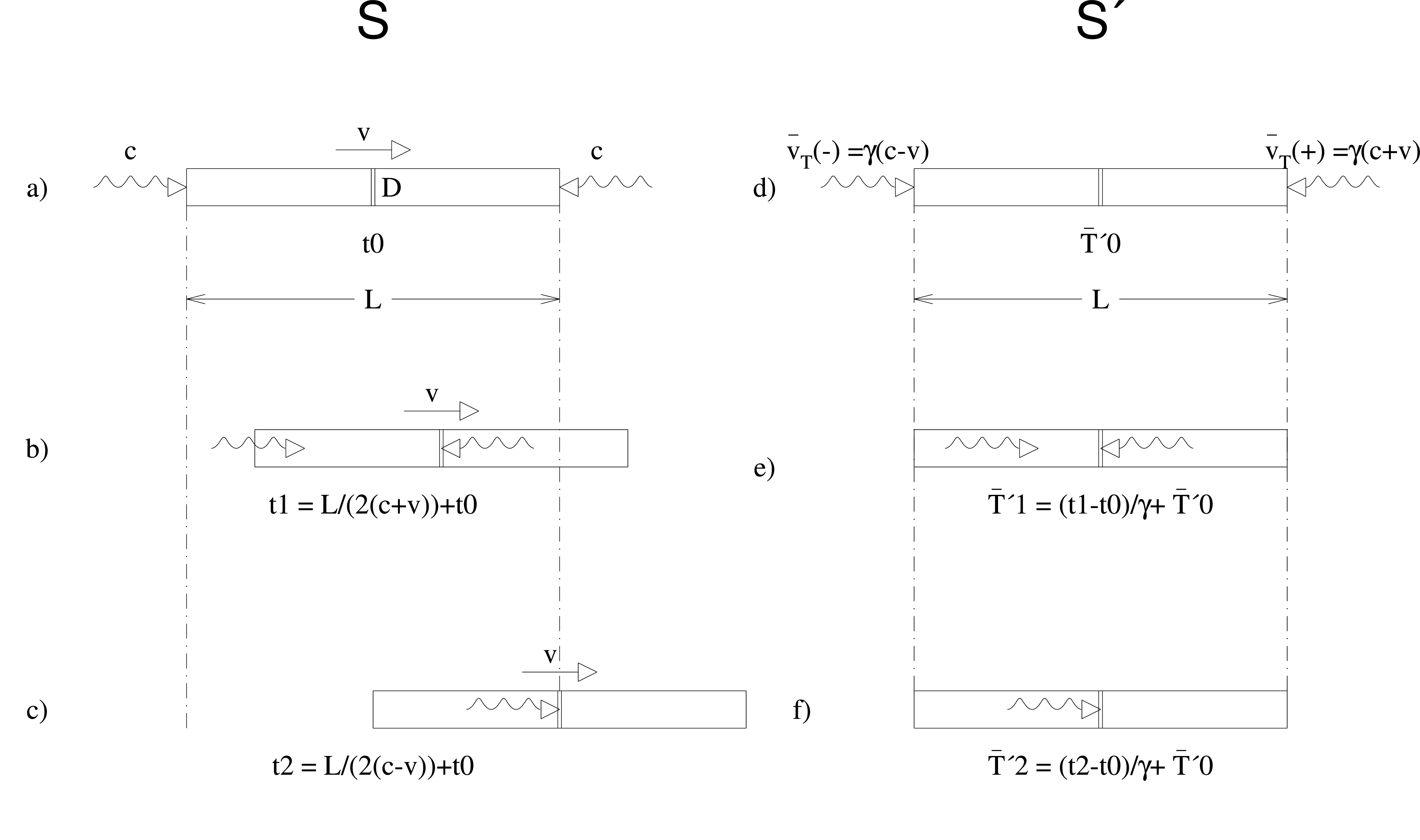}}
\caption{{\em Analysis of Einstein's train-embankment thought experiment.
   Configurations a),b) and c) in the embankment frame (S); c),d) and e) in the train frame (S').
    $v = c/2$, $\gamma = 2/\sqrt{3}$. See text for discussion.}}
\label{fig-fig4}
\end{center}
\end{figure}

       \par This train-embankment thought experiment (TETE) is now analysed in terms of
       the concepts and nomenclature introduced above. The observer O$_T$
       is replaced by a two-sided light detector, D, at the middle of the train. The
       latter moves to the right with speed $v$. The embankment frame, S, is the base and source
        frame of the experiment, the train frame, S', is the travelling, target and subject frame, since
     the aim is to calculate the times of emission of the light signals received by D, i.e. in the
     train frame. At time $t0$ in S (Fig. 4a) light signals moving at speed $c$ in the embankment frame
     are emitted, and move towards D. The light signals are also `travelling objects' in the source frame
     S. The essential input parameters of the problem, $v$ and $c$ are therefore fixed in the frame
     S. In accordance with Eqn(3.18) the length of the train, $L$, is invariant. At time in S
      $t1 = L/[2(c+v)] + t0$ (Fig. 4b) the left-moving light signal strikes D, and at time in S
   $t2 = L/[2(c-v)] + t0$.(Fig. 4c) the right-moving light signal strikes D. The configurations
     in S' corresponding to those in S in Figs. 4a,b,c are shown in  Figs. 4d,e,f respectively.
     The velocity transformation formula (3.31) implies that the speed in S' of the
    right-moving light signal relative to D is $\bar{v}_T(-) = \gamma(c-v)$ while that of the
 left-moving light signal is $\bar{v}_T(+) = \gamma(c+v)$. The pattern of events in S and S' is
    then the same, the only difference being that that the velocities of the light
    signals relative to D are greater in S' by the factor $\gamma$ ---a necessary consequence of
    time dilation and the invariance of length intervals. The left-moving light signal
    is then observed in S' at the time:
   \begin{equation}
      \bar{T}'1 = \frac{L}{2 \gamma(c+v)} +\bar{T}'0 = \frac{t1-t0}{\gamma} + \bar{T}'0 
    \end{equation}
     and the right-moving one at the time:
   \begin{equation}
      \bar{T}'2 = \frac{L}{2 \gamma(c-v)} +\bar{T}'0 = \frac{t2-t0}{\gamma} + \bar{T}'0 
    \end{equation}
     The time dilation  effect for the travelling frame S' is manifest in these equations.
     \par On the assumption that an experimenter analysing the signals recieved by D knows
      the essential parameters of the problem, $L$, $v$, and $c$, the measured times 
     $\bar{T}'1$ and $\bar{T}'2$ in the train frame can be used to decide whether the 
       left and right moving light signals were emitted simultaneously in this frame or not.
        If the right-moving and left-moving signals are emitted at times $\bar{T}'0(-)$ and
        $\bar{T}'0(+)$ respectively then (3.55) and (3.56) are modified to:
   \begin{equation}
      \bar{T}'1 = \frac{L}{2 \gamma(c+v)} +\bar{T}'0(+) = \frac{t1-t0}{\gamma} + \bar{T}'0(+) 
    \end{equation}
     and 
   \begin{equation}
      \bar{T}'2 = \frac{L}{2 \gamma(c-v)} +\bar{T}'0(-) = \frac{t2-t0}{\gamma} + \bar{T}'0(-). 
    \end{equation}
     Subtracting (3.57) from (3.58) and rearranging:
   \begin{equation}
     \bar{T}'2 -  \bar{T}'1 =  \bar{T}'0(-)- \bar{T}'0(+) +\frac{\gamma \beta L}{c}.
   \end{equation}
   The observed time difference $\bar{T}'2 -  \bar{T}'1$ and knowledge of the value
   of $\gamma \beta L/c$ then enables determination of $\bar{T}'0(-)- \bar{T}'0(+)$ so
    the simultaneity of emission of the light signals can be tested. For the event configurations
     shown in Fig. 4 it would be indeed concluded that $\bar{T}'0(-)= \bar{T}'0(+)$, 
      so the emission of the signals is found to be simultaneous in the train frame,
      contrary to Einstein's assertion in Ref.~\cite{EinTETE}. The essential flaw in Einstein's
    argument was the failure to distinguish between the speed of light, relative to some fixed
      object in an inertial frame, and the speed of light relative to some moving object
     in the same frame, which is what is relevant for the analysis of the TETE.
      Einstein's interpretation corresponds to replacing  $\bar{T}'2$ and  $\bar{T}'1$
      by $t2$ and $t1$, so that only events in the embankment frame are 
      considered, and making the replacements, (confusing the speed of light in an inertial frame, with the relative
      speed of light and a moving object in the same inertial frame): $\gamma(c \pm v) \rightarrow c$ in (3.57) and (3.58),
      giving:
    \begin{equation}
       \bar{T}'0(-)- \bar{T}'0(+) = t2  - t1 = \frac{\gamma^2 \beta L}{c}.
    \end{equation}
       This gives Einstein's false conclusion that the light signal emission
       events would be found to be non-simultaneous in the train frame.

    \SECTION{\bf{ Inverse space-time transformations and reciprocal experiments}}      
     In this section the physical concept of {\it reciprocal space-time experiments}, crucial for a correct
      understanding of the `twin paradox' of special relativity will be developed and contrasted with
      the mathematical concepts of a space-time coordinate transformation and its inverse. 
      \par the generic space-time LT is:
       \begin{eqnarray}
       \xr' & = & \gamma(\xr- v \tr), \\
  \tr' & = & \gamma(\tr-\frac{v \xr}{c^2})
       \end{eqnarray}
         while its inverse is    
      \begin{eqnarray}
       \xr & = & \gamma(\xr'+ v \tr'), \\
  \tr & = & \gamma(\tr'+\frac{v \xr'}{c^2}).
       \end{eqnarray}
  Eqns(4.3) and (4.4) are a necessary algebraic consequence of (4.1) and (4.2) and the
  definition $\gamma \equiv 1/\sqrt{1-(v/c)^2}$. It is implicit in these equations
  that clocks in the frame S (coordinates $\xr$, $\tr$) and S'(coordinates $\xr'$, $\tr'$) 
  are synchronised so that $\tr = \tr' = 0$ when $\xr  = \xr' = 0$. Roman symbols are used here and in Section 6
   below to denote generic space and time coordinates, in the absence of the qualifying labels and symbols 
   introduced in the previous section, and additional additive constants necessary to describe correctly
    synchronous clocks for which  $\xr'  \ne  0$~\cite{JHFLLT,JHFUMC,JHFCRCS,JHFACOORD,JHFSTP1}.  
   \par The LT (4.1),(4.2) and its inverse (4.3),(4.4) are now used to discuss in detail a
    particular space-time experiment ---observation of the TD effect--- employing the nomenclature and
    notation introduced in the previous section. Clocks C and C' are introduced at the origins O and O'
    of S and S', respectively and the case where the moving clock C' is viewed by an observer at
   rest in S is first considered. In this case the corresponding reciprocal experiment is one
    in which the  moving clock C is viewed by an observer at rest in S'. The experiment is
     exemplified by the physically interesting case of observation of decay in flight of an
     unstable particle in which the TD relation is used to derive the lifetime $\tb'_D$ of the particle
     in its rest frame. This is done by making use of observations of its flight path $X_B$
     and lifetime $T_B$ in the frame S. In practice, $T_B$ is found from the relation
     $T_B= X_B/v_B$, where $v_B$ is deduced from the relativistic momentum (p) and energy (E)
     of the decay products of the unstable particle via the relation (see Eqn(4.26) below)
     $v_B = p c^2/E$.
      \par The appropriate version of the LT to describe such an experiment
       is the same as Eqns(3.1) and (3.2) above:
           \begin{eqnarray}
             \x'(\C')& = & \gamma_B [X(\C')_B-v_B T(\C)_B] = 0, \\
 \tb'(\C')& = & \gamma_B [T(\C)_B-\frac{v_B X(\C')_B}{c^2}] 
        \end{eqnarray}
         yielding the TD relation
   \begin{equation} 
       T(\C)_B = \gamma_B \tb'(\C').
    \end{equation}
   In the specific case of the decay of an unstable particle, the rest frame decay lifetime 
     is given by (4.7) as:
   \begin{equation} 
    \tb'_D = \frac{T_B}{\gamma_B} = \frac{X_B}{\gamma_B v_B} = \frac{X_B}{\gamma_B}\left(\frac{E}{pc^2}\right)
      = \frac{X_B m}{p}
  \end{equation}
    where $m$ is the mass of the decaying particle. In (4.5)-(4.8) S is the subject, source and base 
     frame while S' is the object, target and travelling frame.
    \par For an observer at rest in S' during the experiment specified by the LT (4.5) and (4.6), 
    $ \tb'(\C') = \Tb'(C')_T$ where $\Tb'(C')_T$ is the time recorded by a clock a rest in his/her own
     (travelling) frame. The TD relation may then be written, for small time increments, as:
      \begin{equation}
       \delta T(C)_B = \gamma_B \delta \Tb'(C')_T.
  \end{equation}
   From the Lorentz invariance of length intervals in the frames S and S', Eqn(3.18), and the fact
   that the direction of motion along the common $x$-$x'$ axis of S' relative to S, in S is 
   opposite to that of S relative to S' in S', it follows that:
        \begin{equation}
     \delta X(C')_B = - \delta \X'(C)_T 
  \end{equation}
    where $\X'(C)_T$ is the coordinate in S' of the origin of S. 
    Differentiating (4.5), using (4.9) and (4.10) and taking the limit of vanishing intervals
    gives the same relation  between derivatives as (3.20):
      \begin{equation}
       \frac{d X(C')_B}{dT(C)_B} \equiv v_B =  -  \frac{d \X'(C)_T} {\gamma_B d \Tb'(C')_T}  \equiv
     \frac{\bar{v}'_T}{\gamma}
    \end{equation}
      so that
     \begin{equation}
   \bar{v}'_T = \gamma_B v_B
 \end{equation}
    which is the same as Eqn(3.21) above.
   \par As shown by the derivation of Eqn(3.37) above, identical predictions are obtained
    if the source and target frames are exchanged so that the LT inverse to (4.5) and (4.6)
    may be used to obtain the TD relation:
       \begin{equation} 
       \Tb(\C)_B = \gamma_B t'(\C').
    \end{equation}
    The TD predictions are therefore invariant with respect to exchange of source and target
    frames, that is, to the use of a transformation or of its inverse.
    However {\it this is not the case when experiments with reciprocal
       configurations are performed,
    i.e. when both subject and object and base and travelling frames are exchanged}.
     This point is crucial for the correct understanding of the `twin paradox' 
     Predictions for an experiment with a configuration reciprocal to the one just described are given by 
      exchange of primed for unprimed quantities in the above equations and reversing
     the signs of all velocity parameters. Hence, in the reciprocal configuration,
     the TD relation (4.7) is replaced by
        \begin{equation}
        T'(C')_B = \gamma'_B \tb(C)
  \end{equation}
         and the velocity relation (4.12) by
       \begin{equation}
   \bar{v}_T = \gamma_B' v'_B.
 \end{equation}
   The relations (4.12) and (4.15) show that the `Reciprocity Principle' (RP)~\cite{BG}, mentioned in the Introduction,
   which has hitherto been assumed to hold in both Galilean and special relativity requires re-interpretation
    in special relativity. The RP states that:
 \par {\bf (RP) If the velocity of an  inertial frame S' relative to another such frame S is
        $\vec{v}$, then the velocity of S relative  to S' is  $-\vec{v}$.} 
    \par  The RP has hitherto, following Einstein in Ref.~\cite{Ein1}, been assumed to describe
        velocity observations in the frames S and S' of a experiment in which events in the 
        two frames are connected by the space-time LT. Such a interpretation is evidently at variance
     with Eqns,(4.12) and (4.15). The corresponding principle in special relativity may be called
     the `Kinematical Reciprocity Principle' (KRP):
 \par {\bf (KRP) The velocity of an inertial frame S' relative to an inertial frame S in
     primary space-time experiment is equal and opposite to the velocity of S relative to S'
    in the reciprocal experiment.} 
    \par Unlike the RP, the KRP describes a relation between the kinematical
     configurations of two physically distinct experiments, not that between observations
     in two different frames in the same experiment, as in (4.12) and (4.15).
   As will be seen below, although the events 
    of a primary experiment are physically distinct (not related by the space-time LT) from those
     of the reciprocal experiment, the base frame kinematical configurations of a primary experiment
     and its reciprocal {\it are} related by
    a {\it kinematical} LT of the relativistic energy and momentum of the objects concerned 
      (see below) or, equivalently, the base frame velocity transformation
     formulas (3.28), (3.32) and (3.33) above.

   The experiments yielding the TD relations (4.7) and (4.14) may be described as having 
    {\it reciprocal configurations}.  
   In the case that the initial conditions of the experiment and the experiment with a 
   reciprocal configuration are the same:
       $v_B = v'_B$ and $\gamma_B  = \gamma'_B$, identical  predictions (after exchange of primed
     and unprimed quantities) are obtained for an experiment and its reciprocal, i.e. the clock
     C' as viewed from S is slowed down by the same ratio in comparison to a clock at rest as when
     C when viewed from S'. Then the two experiments may be called {\it reciprocal} ones.
     It can now be seen that the KRP is actually the {\it definition} of a reciprocal expriment,
     given some primary one, not some relation between observations in S and S' for the primary
     experiment. 
     Conversely, if an experiment and its reciprocal give (after exchange
     of primed and unprimed quantities) identical results then necessarily $v_B = v'_B$. 
       These predictions are summarised in the `measurement reciprocity 
     postulate' (MRP)~\cite{JHFSTP1,JHFHPA}:
        \par {\bf (MRP) Reciprocal space-time measurements of similar rulers and clocks 
            at rest in two different inertial frames S, S', by observers at rest in S', S 
             respectively, yield identical results.}  
     \par As shown in Refs.~\cite{JHFSTP1,JHFHPA} the MRP, together with the requirement that 
      the space-time transformation equations are single-valued functions of their arguments
      is sufficient to derive the LT (4.1)and(4.2) without consideration of any `light signals'
     i.e. independently of Einstein's scond postulate. This type of derivation of the LT
     was first performed by Ignatowsky in 1910~\cite{Ignatowsky}. 
      \par The crucial point for the understanding of the twin paradox is that the experiment
      and its reciprocal are {\it completely independent of each other}. Thus the TD
      relation (4.7) where the clock C' is seen to run slower than C is an experiment
      performed by an observer at rest in S, while the TD relation (4.14) where the clock C is seen
      to run slower than C' is a different experiment performed by an observer at rest in S'.
      Because the experiments are different the question of any contradiction between the
      predictions for them cannot, even in principle, arise. 
     \par In special relativity then, just as in quantum mechanics, the results of an experiment
      depend both on the {\it a priori} initial conditions and how, and by whom, the experiment is performed.
      The predictions of special relativity describe therefore not the properties of some 
      abstract mathematical `space-time' as envisaged by Minkowski~\cite{Minkowski} but the
     expected space-time measurements of some particular observer. In this respect thay are very similar
     to those of linear perspective for the perception of three-dimensional space, which is
     also intrinsically observer-dependent and based on the mathematics of projective geometry. 
      In fact the TD effect corresponds
      to a $\Delta x' = 0$ projection of the LT~\cite{JHFAJP1}.
      \par Without any further discussion, as in the following section, of the differential
       aging implicit in (4.7) and (4.12) on the one hand, or (4.14) and (4.15) on the other,
        it can already be seen that no paradox of self-contradiction can arise by comparison
       of the predictions for an experiment with those of a reciprocal one {\it even if
        the same clocks with the same initial settings are used in both experiments}.
        The physical meaning of $t'(\C')$ in
        (4.13) and $T'(C')_B$ in (4.14), although both are time intervals recorded by the
        same clock, are completely different. Indeed, setting $\tr' = t'(\C') = T'(C')_B$
   and $\tr = \tb(\C) = \Tb(\C)_B$ as in the generic LT equations (4.1) and (4.2) gives, instead
       of (4.13) and (4.14), respectively:
    \begin{eqnarray}
          \tr & = & \gamma_B \tr', \\
     \tr' & = & \gamma'_B \tr.
    \end{eqnarray}
    These equations require either that $\tr =  \tr' = 0$ or that $\gamma_B \gamma'_B = 1$,
     i.e. $\gamma_B = \gamma'_B = 1$, $v_B = v'_B = 0$,
    in contradiction to the assumed initial condition of the experiment  $v_B = v'_B > 0$.
    In order to avoid this antimony it is essential that the base frame and travelling frame
    of both the experiment and its reciprocal be properly specified for the time intervals measured in each
    subject frame. However, as previously pointed out, the predictions for either an experiment
    or its reciprocal are invariant with respect to exchange of source and target frames.
    \par Although an experiment and its reciprocal are completely independent concerning the space-time
        events occuring in either one or the other of them, it is important to note that the
        {\it initial kinematical configurations}\footnote{A `kinematical configuration' in a given 
         inertial frame is specified by the vectorial velocities and spatial separations of physical
          objects in the frame at any instant} of an experiment and its reciprocal {\it are} related
          by a kinematical (momentum-energy) LT\footnote{For a massive physical object an
         energy-momentum LT is equivalent to a LT of the 4-vector velocity of the object. The
        latter is however undefined (all components are infinite) for a massless object, whereas
        the energy-momentum LT is well-defined for both massive and massless objects (See below).}.  Denoting by
      $d\tau({\rm O})$ an element of proper time of the object O considered in Eqns(3.24) and (3.25) above,
      these equations give, on dividing throughout by  $d\tau({\rm O})$ and multiplying throughout by 
          the Newtonian mass $m$ of the object, the LT of its momentum and energy:
         \begin{eqnarray}
        p'({\rm O}) & = & \gamma_B[ p({\rm O})-v_B E({\rm O})/c^2], \\
      E'({\rm O}) & = & \gamma_B[ E({\rm O})-v_B p({\rm O})]
        \end{eqnarray}
          where
     \begin{eqnarray}
 p'({\rm O}) & \equiv & m \frac{dx'({\rm O})_B}{d\tau({\rm O})} = 
        \gamma_{u'_B} m \frac{dx'({\rm O})_B}{d\bar{t}'({\rm O})_B} =  \gamma_{u_B'} u_B' m, \\
     p({\rm O}) & \equiv & m \frac{dX{\rm O})_B}{d\tau({\rm O})} =
       \gamma_{\bar{w}_B} m \frac{dX({\rm O})_B}{d T({\rm O})_B} =\gamma_{\bar{w}_B}  \bar{w}_B m, \\
 E'({\rm O}) & \equiv & m c^2 \frac{d \bar{t}'({\rm O})_B}{d\tau({\rm O})} = 
         \gamma_{u_B'} m c^2, \\ 
 E({\rm O}) & \equiv & m c^2 \frac{d T({\rm O})_B}{d\tau({\rm O})} = 
        \gamma_{\bar{w}_B} m c^2
 \end{eqnarray}
   and $\gamma_{u_B'} \equiv 1/\sqrt{1-(u_B'/c)^2}$. Multiplying (4.18) on both sides by $c^2$ and dividing
     by (4.19) gives:
     \begin{eqnarray}
 \frac{c^2 p'({\rm O})}{ E'({\rm O})} & = & \frac{c^2 p({\rm O})/ E({\rm O})-v_B}{1-v_B p({\rm O})/ E({\rm O})}
    \nonumber \\  
    & = & u_B' = \frac{\bar{w}_B-v_B}{1- \frac{\bar{w}_B v_B}{c^2}}
   \end{eqnarray}
   where the relations: $\bar{w}_B = c^2 p({\rm O})/ E({\rm O})$ following from (4.21) and (4.23) and 
     $u_B' = c^2 p'({\rm O})/ E'({\rm O})$ from  (4.20) and (4.22) have been used. Eqn(4.24)
    is a transposition of the parallel velocity addition formula (3.28). If $-u_B' = v_B = v'_B$ then
      $\bar{w}_B = 0$ so that the object is at rest in S. The kinematical configurations
       of a primary experiment with O at rest in S and its reciprocal where O moves with speed
         $v'_B = v_B$ along the negative $x'$-axis in S' are therefore related by the 
       kinematical LT (4.18) and (4.19). 
     \par For a massive object, the energy-momentum transformation equations (4.18) and (4.19) are 
      equivalent to those:
           \begin{eqnarray}
        U' & = & \gamma_B[U -\frac{v_B}{c} U_0], \\
   U'_0 & = & \gamma_B[U_0 -\frac{v_B}{c}]
\end{eqnarray} 
    of its 4-vector velocity:
       \[  U \equiv (\gamma_u c; \gamma_u \vec{u}) \equiv (U_0; \vec{U}). \]    
    For a massless object (for example a photon) the  4-vector velocity is physically meaningless, as all its
     components are infinite, but the the momentum-energy transformations (4.18) and (4.19)
     and the base frame velocity addition formula (3.28) or (4.24) remain valid. This is because
    both $E = \gamma m c^2$ and $p =  \gamma m v$ remain finite in the limit $m \rightarrow 0$, $v \rightarrow c$
      $\gamma \rightarrow  \infty$ as a consequence of the relation $E^2 = m^2 c^4 +p^2 c^2$~\cite{JHFHPA}.

   \SECTION{\bf{ The twin paradox}}
    Before entering into a more detailed discussion of the differential aging effect (DAE) that is 
      the basis for the correct understanding of the twin paradox a simple space-time experiment is
      described in which the DAE is both manifest and experimentally verified. This is not merely 
      a thought experiment, impossible to realise in practice, but a description of the space-time
      physics that actually governs the use of high energy secondary beams of unstable
      particles. The busy reader need read only this account to be convinced of the physical 
      reality of the DAE effect.\footnote{In spite of having followed the career of an experimental
      particle physicist, it is only very recently that consideration of this
      experiment convinced the present author of the correctness of the DAE that renders younger
     the travelling twin. I had earlier realised that the conventional explanation of the
     twin paradox, to be discussed in the following section, was erroneous, and so had thought it possible,
     that there was no DAE, in spite of the apparent demonstration of the effect in the
     experiment of Hafele and Keating~\cite{HK}. I have since found that similar experiments
     involving charged pions or muons were discussed long ago by Adair~\cite{Adair} and Williams~\cite{Williams}.}  
     \par The `clocks' in the experiment are $\pi^+$ mesons which are created at the same time,
      in the same event by the collision of a high energy (several GeV) extracted proton
      beam with a proton at rest in a hydrogen target. One charged pion, $\pi^+_B$, (analogous to a clock
       at rest in the base (B) frame), produced by fragmentation of the target proton,
       is produced with a very low momentum, loses energy rapidly by ionisation of the liquid
       hydrogen and comes to rest in the target within a short time $\le 1$ns, that may be neglected
       in the following considerations. The other charged pion $\pi^+_T$ ,(analogous to a clock
       at rest in the travelling (T) frame) produced by fragmentation of the beam proton 
         has an energy of 2.8 GeV, corresponding to a speed in the laboratory (base) frame 
        of $0.9988c$ or $\gamma = 20$. It travels almost parallel to the initial proton beam
        over a distance of 100m where 334ns later, in the laboratory frame, it collides elastically, head-on, 
        with another $\pi^+$ at rest in a stopping target (unlikely, but certainly
        physically possible) and itself comes to rest in the laboratory frame. According to the
       nomenclature and notation of Section 3, the subject, source and base frame, S,
       of the experiment is that of the laboratory where  $\pi^+_B$ represents a clock at rest,
       while the object, target and travelling frame, S', is the rest frame of $\pi^+_T$. According to
       the TD relation (4.7) the elapsed time in the rest frame of $\pi^+_T$ is:
        \begin{equation}
           \tb'(\pi^+_T) = \frac{T_B}{\gamma} = \frac{334 {\rm ns}}{20} = 16.7{\rm ns}
       \end{equation}
       Since the mean decay lifetime of a charged pion is 26ns the probability that  $\pi^+_T$
       will survive to be scattered back into the laboratory system is large: $\exp[-16.7/26] = 0.53$.
       It may be said that  $\pi^+_T$ is only `middle aged' at the time it scatters into
       the laboratory frame.  However the probability that  $\pi^+_B$ ---a twin of $\pi^+_T$, born
      at the same instant--- is still in existence $\simeq$ 334 ns after it is created is only
        $\exp[-334/26] = 2.7 \times 10^{-6}$.  If the 26ns mean lifetime of a charged pion 
         is compared to a human lifetime of 70yr, then  $\pi^+_T$ is only 45 years old on arriving 
        in the laboratory frame after its travels, whereas  $\pi^+_B$ will, at this instant,
         have been dead and buried for 827yr!. Since high energy pion beams have been used to perform
         experiments hundreds of metres from their production point in particle physics laboratories
        all over the world there can be no doubt of the correctness of the above account, i.e.
       of the reality of the DAE. Indeed, the TD effect has been quantitatively confirmed by observations
       of the decay of beam pions~\cite{TDPIDEC}.

       \par To give more insight into the DAE and its relation to discussions of the twin paradox that
          have appeared in text books and the pedagogical literature, a more elaborate thought 
           experiment, inspired by the simple one involving a charged pion just described, is now
          presented. To understand the role (if any) of acceleration in the DAE and the twin
          paradox it is convenient to introduce macroscopic objects that can, by analogy with the charged
        pion in the above experiment, undergo head-on elastic collisions and so be transfered, with negligible
         time delay, between two inertial frames. These objects may be called SBs for `Space Billiards'. 
         Like ideal billiard balls undergoing perfectly elastic collisions, they are reduced to a 
         state of rest when colliding frontally with an identical object initially at rest. The SBs
         are assumed to move freely in field-free space, without rotation, according to Newton's
          First Law of motion. 
         \par It will also be found convenient, following the original suggestion of Langevin
           ~\cite{Langevin}, to dope certain of the SBs with a radioactive substance with a suitably
         chosen mean decay lifetime, $\tau_D$. The age, $a_{{\rm SB}}$ of such an SB is then in 
           one-to-one correspondence with the number $N$ of undecayed radiactive nuclei remaining
         at any instant, according to the relation, expressing the exponential decay law:
          \begin{equation}
          a_{{\rm SB}} = \tau_D\ln\left( \frac{N_0}{N}\right)
          \end{equation}
         where $N_0$ is the initial number of radioacive nuclei. The age difference of two such
           identically-prepared SBs is then given by the ratio of their activities, $A$, (number of
             radioactive decays per unit time) as measured at the same instant by two identical
           detectors:
      \begin{equation}
          a_1 - a_2 = \tau_D\ln\left( \frac{A_2}{A_1}\right)
          \end{equation}
          \par In the proposed experiment, identical doped SBs are prepared at rest in the subject,
            source, and base frame, S. By elastic collisions with other, identical, moving, SBs, they
          are then projected into the object, target and travelling frame S'. Finally by further
          elastic collisions with identical SBs at rest in S they are projected back into
               this frame, and their ages are compared to those of a similar SB that remained
            at rest in S during their journey. Thus, instead of the round-trip scenario of the
          classical twin paradox, only the outward journey is considered. Because of the symmetry
         of outward and return journeys, both with respect to accelerations, decelerations,
         and times of passage, this simplification does not affect in any way the
         connection between the DAE observed and the resolution of the round-trip twin paradox.
         Also, the analysis of the acceleration-free sections of the journey is identical to that
         in the conventional twin paradox experiment where the travelling twin may be accelerated 
        and decelerated during non-neglible proper time intervals. Since the stay-at-home twin (see below)
        undergoes identical acceleration and deceleration as the travelling twin, but has a negligibly
        short interval of uniform motion, the effect of acceleration and deceleration
       cancels completely in the calculation of the DAE, and so cannot be responsible for it.

\begin{figure}[htbp]
\begin{center}\hspace*{-0.5cm}\mbox{
\epsfysize14.0cm\epsffile{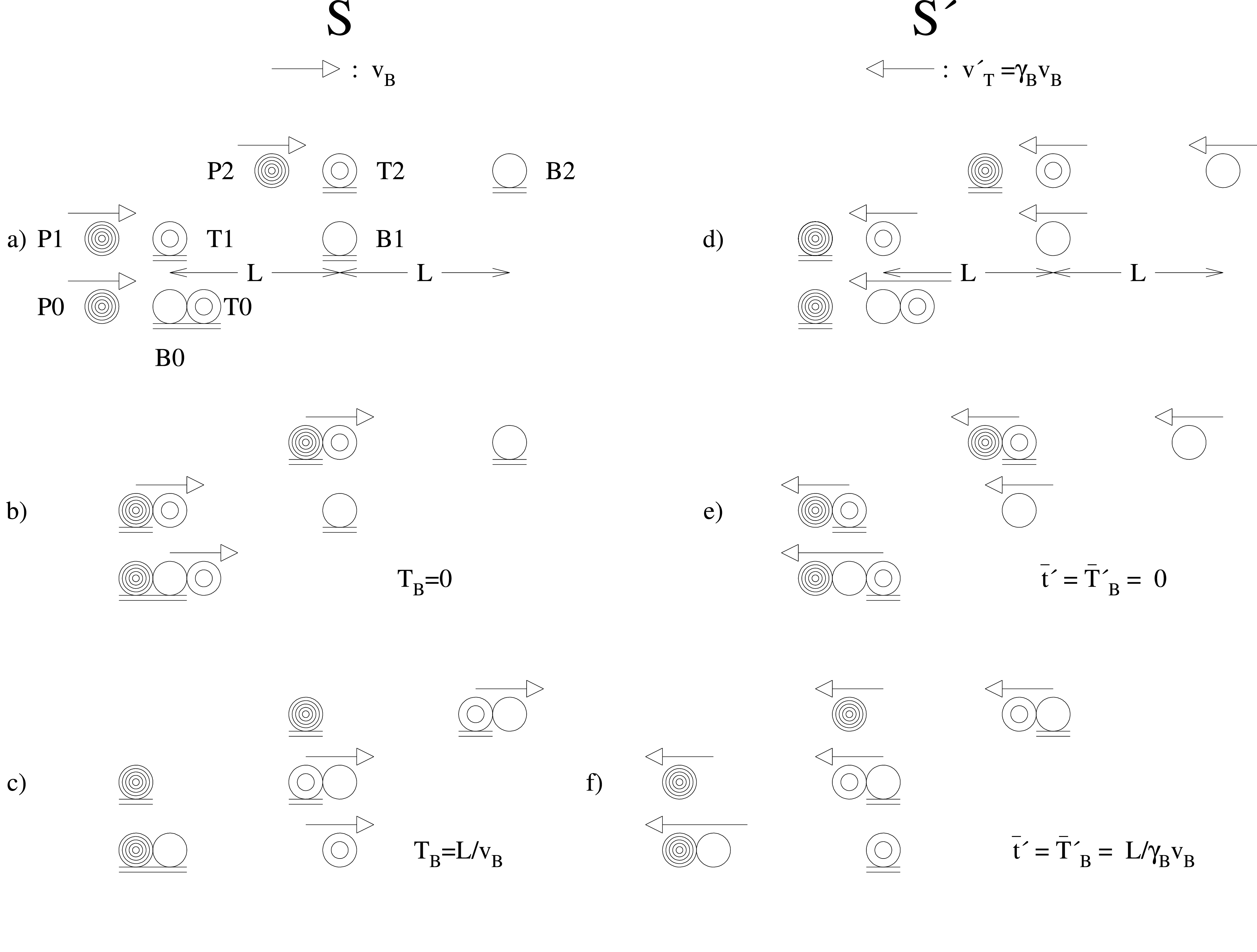}}
\caption{ {\em The space billiard thought experiment. a) initially, the
    SBs B0,B1,B2,T0,T1 and T2 are at rest in the base frame S at the
    positions shown. b) At time T$_B = 0$, T0, T1 and T2 are simultaneously set in
    motion by head-on elastic collisions with P0, P1 and P2. SBs at
    rest are marked by double horizontal lines and SBs in motion by
    arrows. c) at time  T$_B = L/v_B$ T1 and T2 are brought to rest by
    collisions with B1 and B2. Event configurations in the travelling 
     frame S' corresponding to a), b) and c) in S are shown in
     d), e) and f) respectively.}}
\label{fig-fig5}
\end{center}
\end{figure}

          \par As shown in Fig. 5a, initially, six SBs B0,B1,B2,T0,T1 and T2, each similarly, and simultaneously,
           doped with a radioactive substance. and so with the same age according to Eq. (5.2),
       are at rest in the frame S. The SBs B1 and B2 are
           separated from T1 and T2 respectively by the same distance, $L$, and B1 and T2 have the
         same $x$-coordinate. The SB B0, which will constitute the clock at rest in the
          frame S, (the analogue of the stay-at-home twin) is separated from T0 by a short
            distance (not shown) that is much less than $L$, but sufficiently large that B0 attains
           a uniform speed after its collision (see below) with P0. The SBs T0, T1 and T2 are
            set in motion
             by synchronous frontal elastic collisions with  P0, P1 and P2 which each move
            with uniform speed $v_B$ parallel to the $x$-axis (see Fig. 5b). Whereas  T1 and T2
             (the analogues of the travelling twin) are set in motion directly by collisions
              with P1 and P2, P0 first collides with B0, that after a negligibly small displacement
           collides with  T0, setting it in motion and itself coming to rest. Since T1 and T2 are
            later brought to rest in S by elastic collisions with B1 and B2 respectively, identical
              accelerations and decelerations are undergone by B0, T1 and T2 so completely
            cancelling their contributions to the DAE when comparing the age of B0 with that
            of T1 or T2. The collisions of T1 with B1 and T2 with B2 occur, as shown Fig. 5c,
             at $T_B = L/v_B$ in the frame S. The S' frame configurations corresponding to
              Figs. 5a, 5b and 5c, calculated using Eqns. (3.3),(3.9) and (3.21) are shown
              in  Figs. 5d, 5e and 5f, respectively. The spatial configurations in Figs. 5d, 5e and 5f
           are identical to those in Figs. 5a, 5b and 5c respectively. However, after the collisions
            with P0, P1 and P2, B0, B1 and B2 all move to the left with the speed $v'_T = \gamma_B v_B$. 
            It can be seen that the collisions of T1 with B1 and of T2 with B2 that project T1 and T2
              back into the frame S, where their ages may be compared with that of B0, occur
           simultaneously in both S and S consistent with Eqs. (3.9)-(3.11) above.
  \newpage
             \par If now the ages of the travelling SBs T1 and T2 are compared with that of
               the stay-at-home SB T0, by measuring their activities in similar detectors
              at some later fixed time, special relativity predicts the DAE:
        \begin{eqnarray}
          a({\rm B0}) -  a({\rm T1}) & = &  a({\rm B0}) -  t({\rm T2})
          = \frac{L}{v_B}\left(1-\frac{1}{\gamma_B}\right)  \nonumber \\
         & = & \tau_D\ln\left( \frac{A({\rm T1})}{A({\rm B0})}\right)
         = \tau_D\ln\left( \frac{A({\rm T2})}{A({\rm B0})}\right).
          \end{eqnarray}
   \par Note that the reduced the age increments of T1 and T2 in the frame S', as compared to that
     of B0 in the frame S, are
       due to the increase of the apparent velocity of T1 relative to B1 and T2
        relative to B2 in the frame S': $v'_T = \gamma_B v_B$, as compared to $v_B$ in the
          frame S, the journey being of equal length in the two frames.
        In the hitherto conventional interpretation, to
    be discussed below, the relative velocity is assumed to be $v_B$ in both S and S',
    and the spatial separations in the frame S' are scaled down by a factor $1/\gamma_B$, so that
     the reduced age increments of T1 and T2 in S' are explained by 'length contraction' --- a shorter
     journey than in S at the same speed. 
      Thus the solution (a) corresponding
       to measurement of the length of a train, as discussed in Section 2, is adopted, rather
     than solution (b) corresponding to the configurations shown in Fig. 5. It will be shown that
     the conventional type (a) interpretation of the SB thought experiment, arises from a misinterpretation
     of the space-time LT that implies a breakdown of translational invariance ---an unphysical dependence
       of predictions for times of events in the frame S' on an arbitary choice of spatial coordinate
      system--- as well as coordinate-system-dependent and manifestly unphysical `rapid' and `reverse' aging
      effects. 

\begin{figure}[htbp]
\begin{center}\hspace*{-0.5cm}\mbox{
\epsfysize14.0cm\epsffile{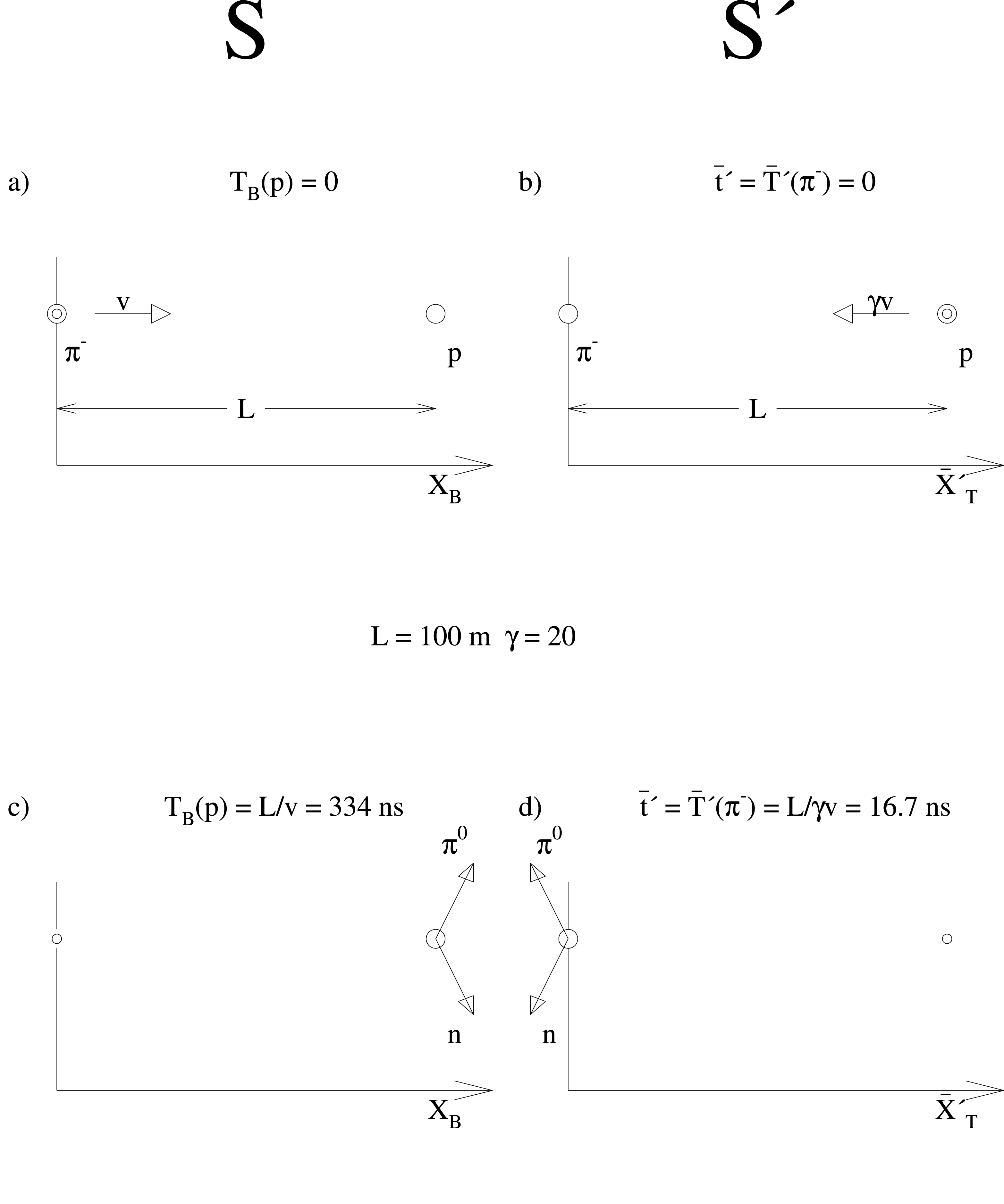}}
\caption{ {\em A $\pi^-$ with $\gamma = 20$ in the base frame S and initially at 100m from
 a proton [a)], collides with  it [c)], and undergoes the charge-exchange reaction
  $\pi^- p \rightarrow n \pi^0$. b) and d) show the corresponding event configurations
   in the travelling frame, S', (the rest frame of the $\pi^-$). See text for discussion.}}
\label{fig-fig6}
\end{center}
\end{figure}

\begin{figure}[htbp]
\begin{center}\hspace*{-0.5cm}\mbox{
\epsfysize15.0cm\epsffile{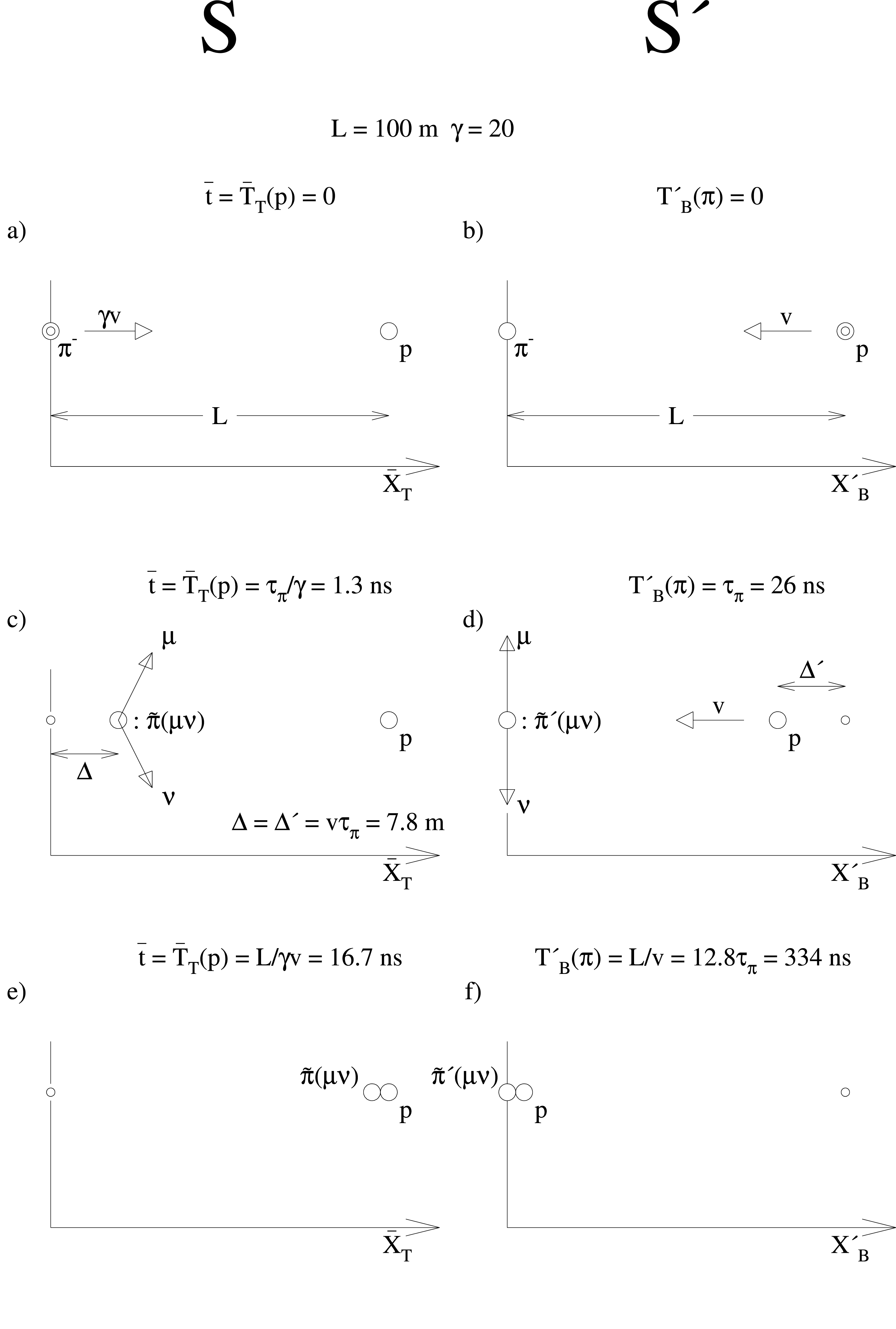}}
\caption{ {\em The experiment reciprocal to the one shown in Fig. 6. b) A proton with $\gamma = 20$ 
  in the base frame S', and 100m from it, is directed towards a  $\pi^-$ at rest in this frame.
   d) At time $T'_B(\pi) = \tau_{\pi} = 26$ns the  $\pi^-$ decays. f) At $T'_B(\pi) = 334$ns
     the $p$ is coincident with, $\tilde{\pi}(\mu \nu)$, the centre of mass the  $\pi^-$
      decay products. The corresponding event configurations in the travelling frame
      (the rest frame of the proton) are shown in a), c) and e). See text for discussion.}}
\label{fig-fig7}
\end{center}
\end{figure}

      \par To further illustrate the crucial importance of the concepts of base and travelling frames, and
       the independent nature of a primary space-time experiment and its reciprocal, a variation of the
        charged pion thought experiment mentioned above is now considered.
        A negative pion $\pi^-_T$ is produced in a similar
           manner to the $\pi^+_T$ of the above experiment. All other physical parameters remain unchanged, 
         but instead of scattering on another charged pion at rest,  $\pi^-_T$ undergoes the 
          charge-exchange reaction $\pi^-_T p \rightarrow \pi^0 n$. Observers in all inertial
           frames must agree on the occurence of this interaction. However, as will now be shown,
        an interaction of $\pi^-_T$ can occur, with an appreciable probability, only in the primary
        experiment as specified above, not in the reciprocal one. The primary experiment with
       the laboratory frame, S, as subject, source and base frame and the rest frame of $\pi^-_T$, S',
       as object, target and travelling frame is show in Fig. 6, with configurations in
       the frame S on the left and in S' on the right. The time interval between the production
        and interaction events is 334ns in S and 16.7ns in S'. As the mean lifetime is
        the same for positive and negative pions, there
       is a 53\% probability that $\pi^-_T$ remains undecayed and is able to initiate the charge
       exchange reaction. The reciprocal experiment is shown in Fig. 7. In this case, S',
       the charged pion rest frame is the subject, source and base frame while the laboratory frame S
       (the proton rest frame) is object, target and travelling frame. In this case the proton
       with which  $\pi^-_T$ can potentially interact is the travelling object. In spite of
        the kinematical and spatial similarity of the  initial states in the primary and reciprocal experiments
        it is essentially impossible that the charge exchange reaction can occur in the
        reciprocal experiment. Typically the pion will decay after a time $\tau_{\pi} = 26$ns,
        (Fig. 7d). In the travelling frame (proton rest frame) such a decay
        occurs after only 1.3ns when the pion is 7.8m from its production point.  Denoting
        by $\tilde{\pi}(\mu \nu)$ the physical system comprised of the pion decay products,
        it is seen in Fig. 5f that the proton is in coincidence with  $\tilde{\pi}(\mu \nu)$
        in S' after a flight tine of 334ns or 12.8$\tau_{\pi}$. The probabilty that the proton
        arrives at the charged pion before it decays, and so can interact with it, is then only
        2.5$\times 10^{-6}$. It can be seen from Figs. 6 and 7 that the possible outcomes
         of two similar experiments, with initial states differing only by
         a Lorentz boost, depends crucially on the choice
        of base and travelling frames, i.e. in which inertial frame the initial conditions
        of the experiment are defined. For example, the potential experience of an observer in the proton
        rest frame is completely different if it is the base frame as in Fig. 6, or the travelling
        frame as in Fig. 7. 

   \SECTION{\bf{ The standard interpretation of the space billiard experiment: with 
           `relativity of simultaneity' and  `length contraction'}} 
       The interpretation of the space billiard thought experiment presented in the previous section
        is summarised in Fig. 8 where, for simplicity, only T1,T1,B1 and B2 are shown. The encounters
         T1,B1 and T2,B2 are simultaneous in both S and S'. The DAE is a consequence of the Lorentz
        invariance of spatial intervals and the relativistic reciprocity relation (3.21) which states that
         the magnitude of the velocity of B1,B2 relative to T1,T2 in the frame S' is a factor $\gamma_B$
        larger than that of  T1,T2  relative to B1,B2 in the frame S. 

\begin{figure}[htbp]
\begin{center}\hspace*{-0.5cm}\mbox{
\epsfysize14.0cm\epsffile{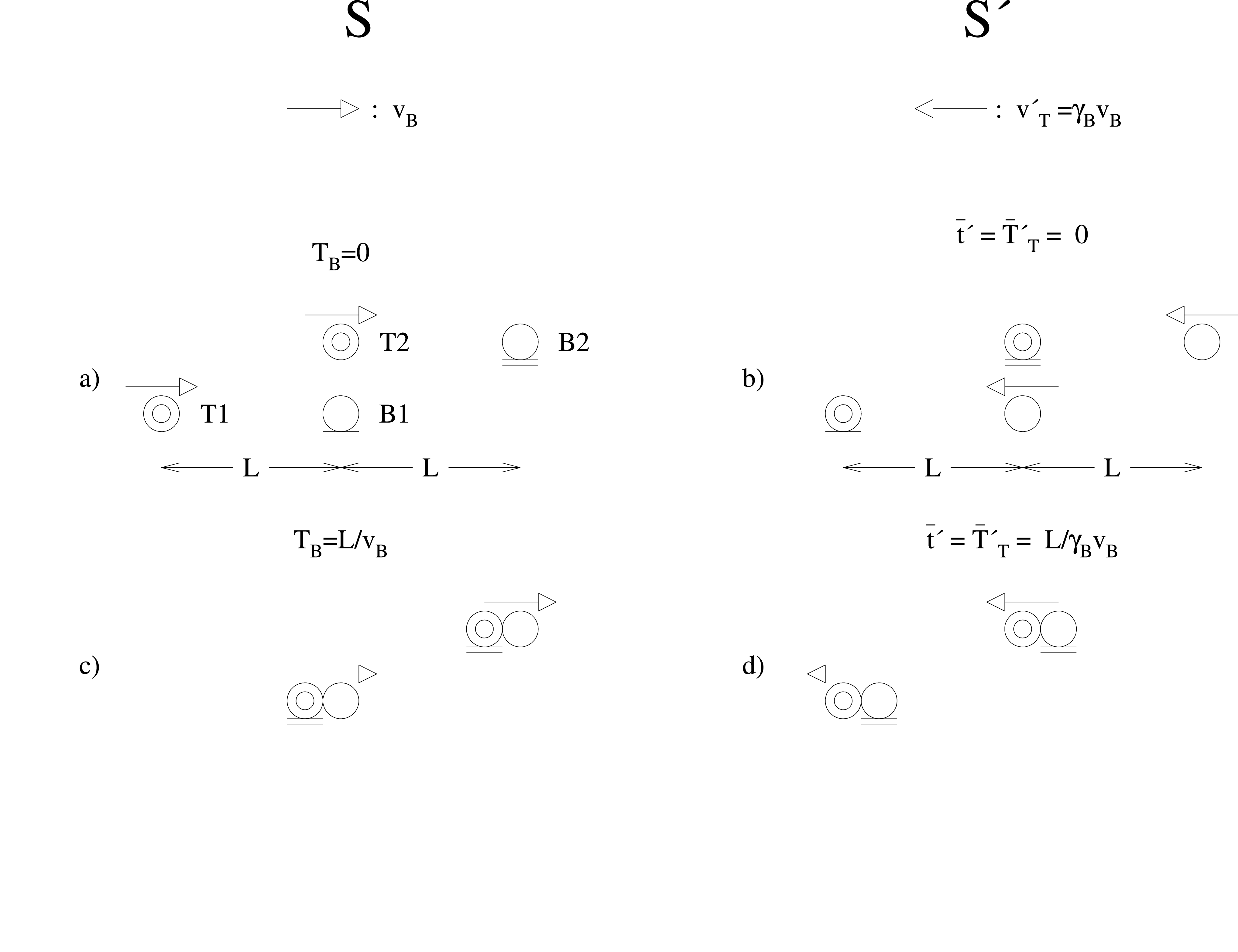}}
\caption{{\em Simplified version of Fig.3 for comparison with the calculation using the generic LT
  (4.1) and (4.2) shown in Figs.8-10. a) T1 and T2 are set in motion, in the base frame S, by
    P1 and P2 (not shown).  b) T1 and T2 are brought to rest in S by collisions with B1 and B2 respectively.
  c) and d) show the corresponding configurations in the travelling frame, S', (rest frame of T1 and T2)
 See text for discussion.}}
\label{fig-fig8}
\end{center}
\end{figure}

          \par In the standard interpretation based on the generic LT (4.1) and (4.2) the space-time 
             configurations in the frame S are identical to those shown in Fig. 8a and  Fig. 8b.
            However, the space-time configurations in S' are found to be much more complicated than
          those shown in Fig. 8c and  Fig. 8d. The positions and times of the SBs in S', for the case 
            where the origin of S is initially at B1 and that of S' initially at T1, calculated with the
           aid of (4.1) and (4.2), are presented in Table 1. These calculated S' frame times are shown on circular
          analogue clocks in correspondence with the positions of the different SBs at $t  = 0$,
          when T1 and T2 have just been set in motion in Fig. 9b. Also shown for each SB on analogue
         clocks with square dials  are the ages of the SBs, corresponding to a zero reading of the
         square clocks at $t = 0$. This age is essentially the same as that at time $t = -\delta$,
       where $\delta \ll L/v$, as in Fig. 9a where all the SBs shown  are still at rest in  S.
       The DAE is given by comparing the settings of these age clocks at $\tr = 0$ in Fig. 9b with those 
       at time $\tr = L/v$ in Fig. 9d where T1 and T2 are again at rest in S. The calculated S' frame times
      and the ages of the SBs corresponding to their positions at the time $t = L/v-\delta$, just 
       before T1 and T2 come to rest in S are shown in Fig. 9c.  
     In drawing Fig. 9 and the subsequent figures in this section it is assumed
    that $\beta = v/c = \sqrt{3}/2$, $\gamma = 2$.
     The times of passage $L/v = 2L/\sqrt{3}c$ of T1 and T2 in S correspond to a rotation of sixty degrees
 of the hands of the clocks. Inspection of Fig. 9 shows that, apart from T2, situated at the origin of S',
  indicated by the large inclined arrow, and B1 at $t = 0$, there is no correspondence
   between the ages of the SBs as given by the
   TD effect, and the clock settings in S' as calculated with the generic LT (4.1) and (4.2).
   \par The pattern of 
   collision events in S' corresponding to the S' frame positions and times presented in Table 1, is shown
   in Fig. 10. At $t'= -\gamma \beta L/c$, (Fig. 10a) before any collisions have taken place, and where
   all objects
   move to the left with speed $v$, the separations of T1 and B1 and T2 and B2 are found to be
   `length contracted' by the factor $1/\gamma$. At $t'=0$ (Fig. 10b) P1 (not shown) collides with T2 and brings
   it to rest in the frame S'. At time $t´ = L/\gamma v$ (Fig. 10c) B2 collides with T2, projecting it back
   into the frame S, and itself comes to rest in S'. The age-increment of T2 while it is at rest in S' is
   therefore $L/\gamma v$, consistent with the TD effect.
   The smaller time interval experienced by T2 in the frame S' is explained here as a 
   consequence of the `length contraction' of the separation of T2 and B2 in this frame, as compared to
   their separation, $L$, in the frame S, their relative velocity being the same in the two frames.
    At $t'= \gamma \beta L/c$ (Fig. 10d) P1 (not shown) collides with T1 setting it into motion in S and 
    bringing it to rest in S'. At $t'= \gamma L/v$ (Fig. 10e) B1 collides with T1, projecting it back into
    the frame S. The elapsed time in S' for which T1 is at rest in this frame:
      $\gamma L/v-\gamma \beta L/c = L/\gamma v$ gives the age increment of T1 in S', again consistent
    with the TD formula. It may be remarked  that the collisions between B2 and T2 (at  $t´ = L/\gamma v$)
      and B1 and T1 ($t'= \gamma L/v$) that are simultaneous in the frame S are separated by the time
     interval $\gamma \beta^2 L/v$ in S', so that in this case there is an infinite DAE when comparing 
     the time interval between the collisions as experienced by observers at rest in S and S'. In contrast, in
      Fig. 8d, the collisions of T1 with B1 and T2 with B2 are simultaneous and only the TD effect is
     experienced by observers in S and S'.
\begin{figure}[htbp]
\begin{center}\hspace*{-0.5cm}\mbox{
\epsfysize15.0cm\epsffile{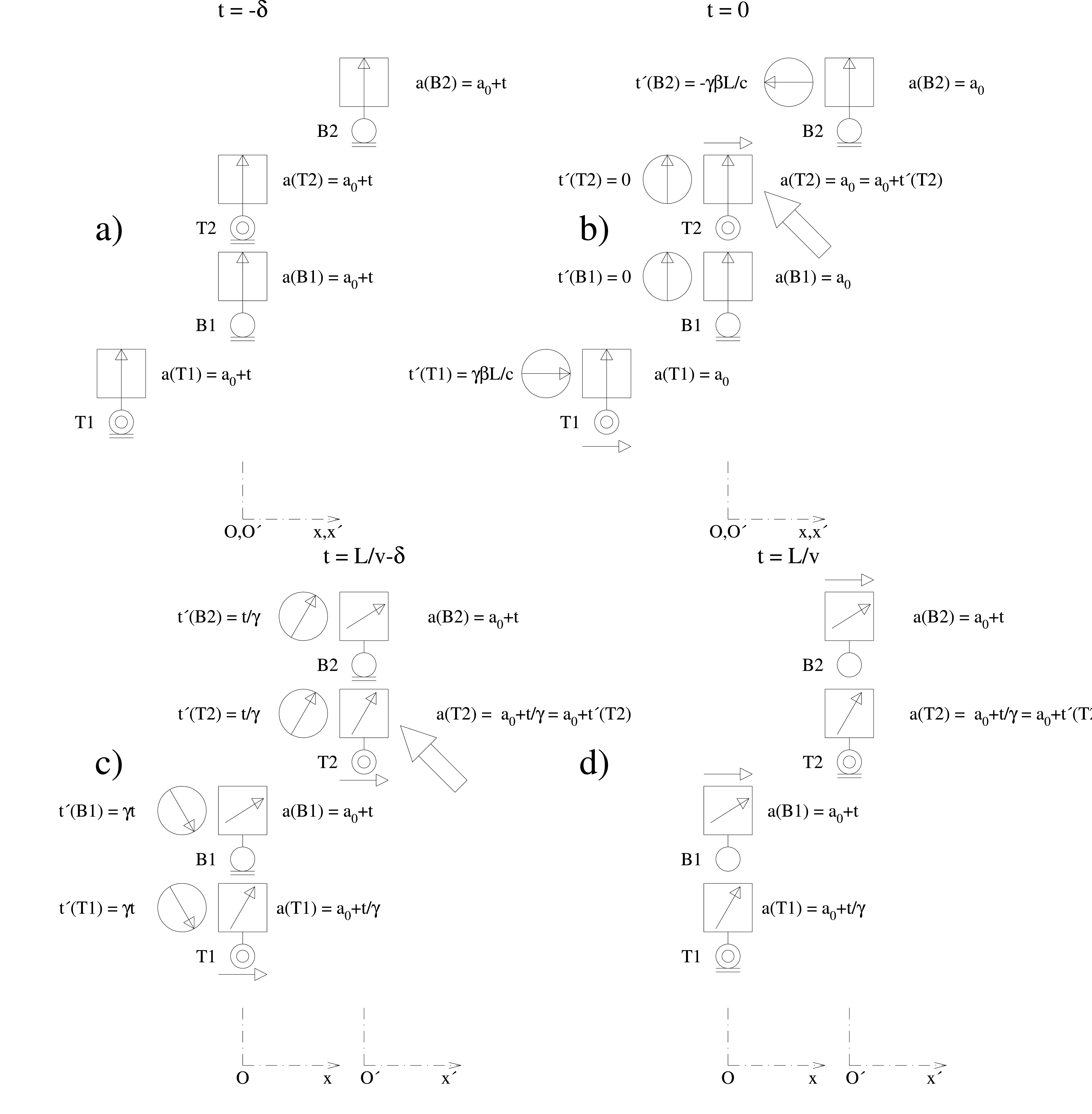}}
\caption{ {\em Analysis of the space billiard thought experiment using the generic LT
 (4.1) and (4.2). Event configurations in the base frame S are shown. The ages of the SBs are shown
   on the analogue clocks with square dials. The analogue clocks with round dials
  show S' (travelling) frame times as calculated using (4.1) and (4.2). a) $t = -\delta$; all SBs
   are at rest and have age $a_0-\delta$. b) T1 and  T2 are set in motion by collisions
   with P1 and P2 (not shown). c) $t = L/v -\delta$; ages and times just before collisions 
    with B1 and B2. d) $t = L/v$; T1 and T2 are brought to rest by collisions with B1 and B2
    respectively. The coordinate origins in S[S'] are aligned with B1[T2].
   $v = (\sqrt{3}/2)c$, $\gamma = 2$. See text for
    discussion.}}
\label{fig-fig9}
\end{center}
\end{figure}

\begin{figure}[htbp]
\begin{center}\hspace*{-0.5cm}\mbox{
\epsfysize15.0cm\epsffile{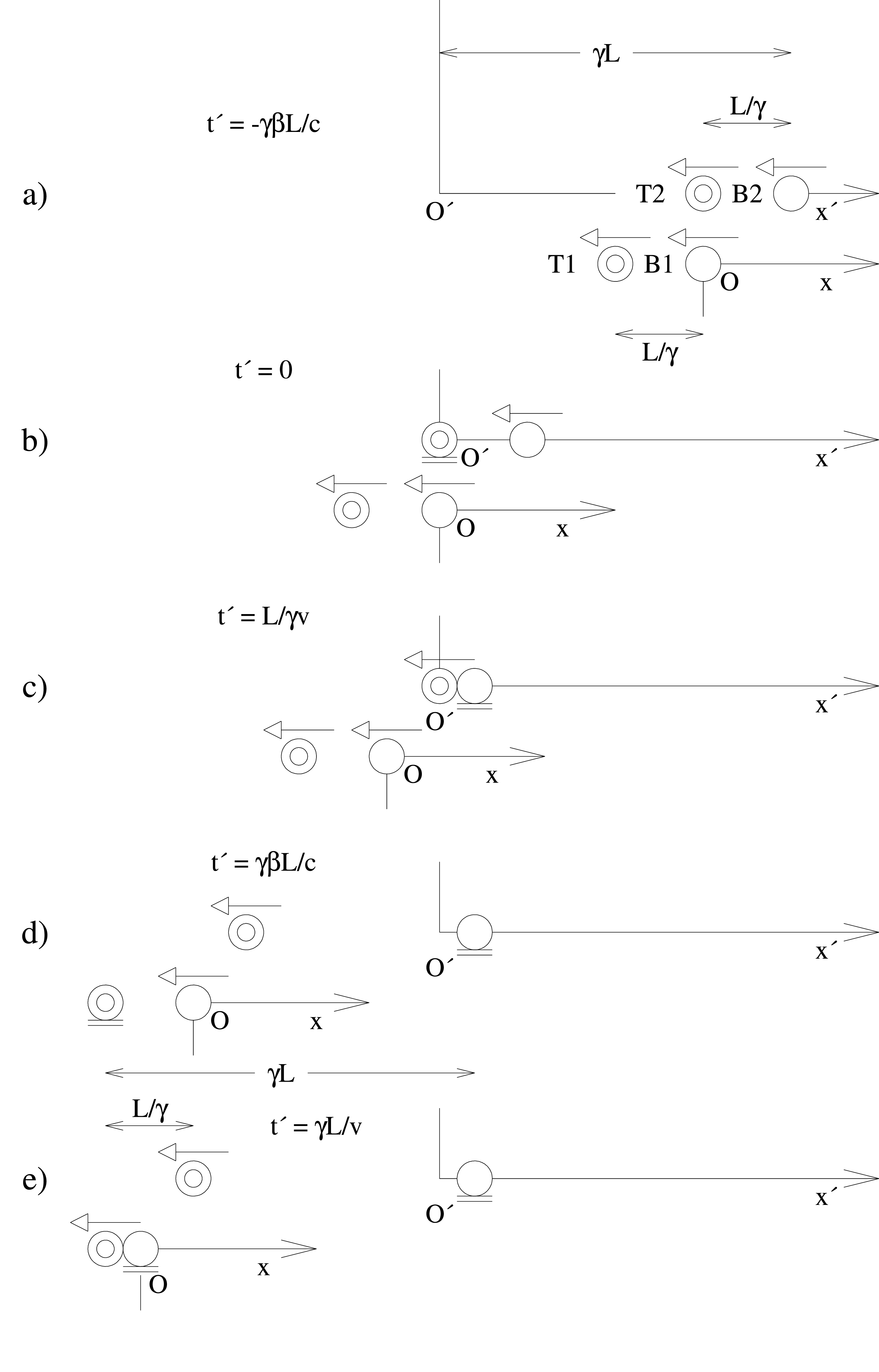}}
\caption{ {\em Event configurations in the travelling frame S' as calculated using (4.1) and (4.2),
 corresponding to the sequence of S (base frame) events shown in Fig. 9. $v = (\sqrt{3}/2)c$, $\gamma = 2$.
 See text for discussion.}}
\label{fig-fig10}
\end{center}
\end{figure}
 
  \begin{table}
\begin{center}
\begin{tabular}{|c||cc|cc|cc|cc|} \hline
\multicolumn{1}{|c||}{ }& \multicolumn{2}{|c}{T1} &\multicolumn{2}{|c}{T2}
&\multicolumn{2}{|c}{B1}&\multicolumn{2}{|c|}{B2} \\ \cline{2-9}
$\tr$ & $\xr'$ & $\tr'$ & $\xr'$ & $\tr'$ & $\xr'$ & $\tr'$ &  $\xr'$ & $\tr'$ \\  \hline \hline
$0$ & $-\gamma L$ & $\gamma \beta L/c$ & $0$ &  $0$ & $0$ &  $0$ & $\gamma L$ & $-\gamma \beta L/c$ \\
 $L/v$ & $-\gamma L$ & $\gamma L/v$ & $0$ &  $L/\gamma v$ & $-\gamma L$ & $\gamma L/v$ & $0$ &  $L/\gamma v$ \\
 \hline 
\end{tabular} 
\caption[]{{\em Space and time coordinates of the SBs in the frame S' as
 calculated using the generic LT (4.1) and (4.2).
 The coordinate origins in S [S'] are aligned with B1 [T2].}}     
\end{center}
\end{table}

\begin{table}
\begin{center}
\begin{tabular}{|c||cc|cc|cc|cc|} \hline
\multicolumn{1}{|c||}{ }& \multicolumn{2}{|c}{T1} &\multicolumn{2}{|c}{T2}
&\multicolumn{2}{|c}{B1}&\multicolumn{2}{|c|}{B2} \\ \cline{2-9}
$\tr$ & $\xr'$ & $\tr'$ & $\xr'$ & $\tr'$ & $\xr'$ & $\tr'$ &  $\xr'$ & $\tr'$ \\  \hline \hline
$0$ & $0$ & $0$ & $\gamma  L$ &  $-\gamma \beta L/c$ &  $\gamma L$ &  $-\gamma \beta L/c$ & 
 $2 \gamma  L$ & $-2\gamma \beta L/c$ \\
$L/v$ & $0$ & $L/\gamma v$ & $\gamma L$  & $-(\gamma^2-2)L/(\gamma v)$ & $0$ & $L/\gamma v$ & $\gamma L$ & 
 $-(\gamma^2-2)L/(\gamma v)$  \\
 \hline 
\end{tabular} 
\caption[]{{\em  Space and time coordinates of the SBs in the frame S' as
 calculated using the generic LT (4.1) and (4.2). B1 is at $x = L$ in S and the
 origin of S' is aligned   with  T1. }}     
\end{center}
\end{table}

      \par With a different choice of spatial coordinate origins ---O' at T1 and O such that $x = L$ for B1---
      the different set of transformed positions and times presented in Table 2 is obtained by the
     use of the generic LT (4.1) and (4.2). The times are displayed in Fig. 11, which is similar to Fig. 9.
 In this case (indicated by the large inclined arrows) the calculated S' frame clock settings are equal to the 
   age of the moving SB only for T1, which is again situated at the origin of S'. The corresponding
   sequence of collision events is shown in Fig. 12. The same pattern of events occurs as in 
     Fig. 10, but shifted backward in time by $\gamma \beta L/c$. This implies that an identical
    set of events in the frame S, when transformed into the frame S' using the generic LT
    (4.1) and (4.2) leads to a different sequence of events (events with different
      times) when different spatial coordinate
    systems are chosen. There is therefore a manifest breakdown of translational invariance, which
    requires that physical predictions in isotropic and homogeneous space must be independent
    of the choice of spatial coordinate system. Indeed, for example, the entries in Table 1 for T1
   at the time $t = L/v$ are in direct contradiction with the translationally-invariant
    TD relation (3.8) that is valid for an arbitary value of $L'$. In this case, where $\xr({\rm T1})= 0$,        ,
     Eq. (4.2) gives:
    \begin{equation}
   \tr'({\rm T1})= \gamma[\tr({\rm T1})-\frac{v \xr({\rm T1})}{c^2}] = \frac{\gamma L}{v}.
    \end{equation}
    To be compared with correct, translationally invariant, formula (3.8) which gives instead for this case,
     \begin{equation}
  a({\rm T1})-a_0 = \tr'({\rm T1})= \frac{\tr({\rm T1})}{\gamma}  = \frac{ L}{\gamma v}.
    \end{equation}         
     Clearly, the formula (6.1) is inconsistent with the TD effect of (6.2) and with the DAE
    that must be the same for T1 and T2 that undergo identical motions in the frame S. In contrast, 
    applying Eq. (4.2) to T2, for which $\xr({\rm T2})= L$ at $t = L/v$ gives:
     \begin{equation}
   \tr'({\rm T2})= \gamma[\tr({\rm T2})-\frac{v \xr({\rm T2})}{c^2}] = \frac{L}{\gamma v}
    \end{equation}
   in agreement with the TD relation for T2:
      \begin{equation}
   a({\rm T2})-a_0 = \tr'({\rm T2})= \frac{\tr({\rm T2})}{\gamma}  = \frac{ L}{\gamma v}.
    \end{equation}

\begin{figure}[htbp]
\begin{center}\hspace*{-0.5cm}\mbox{
\epsfysize15.0cm\epsffile{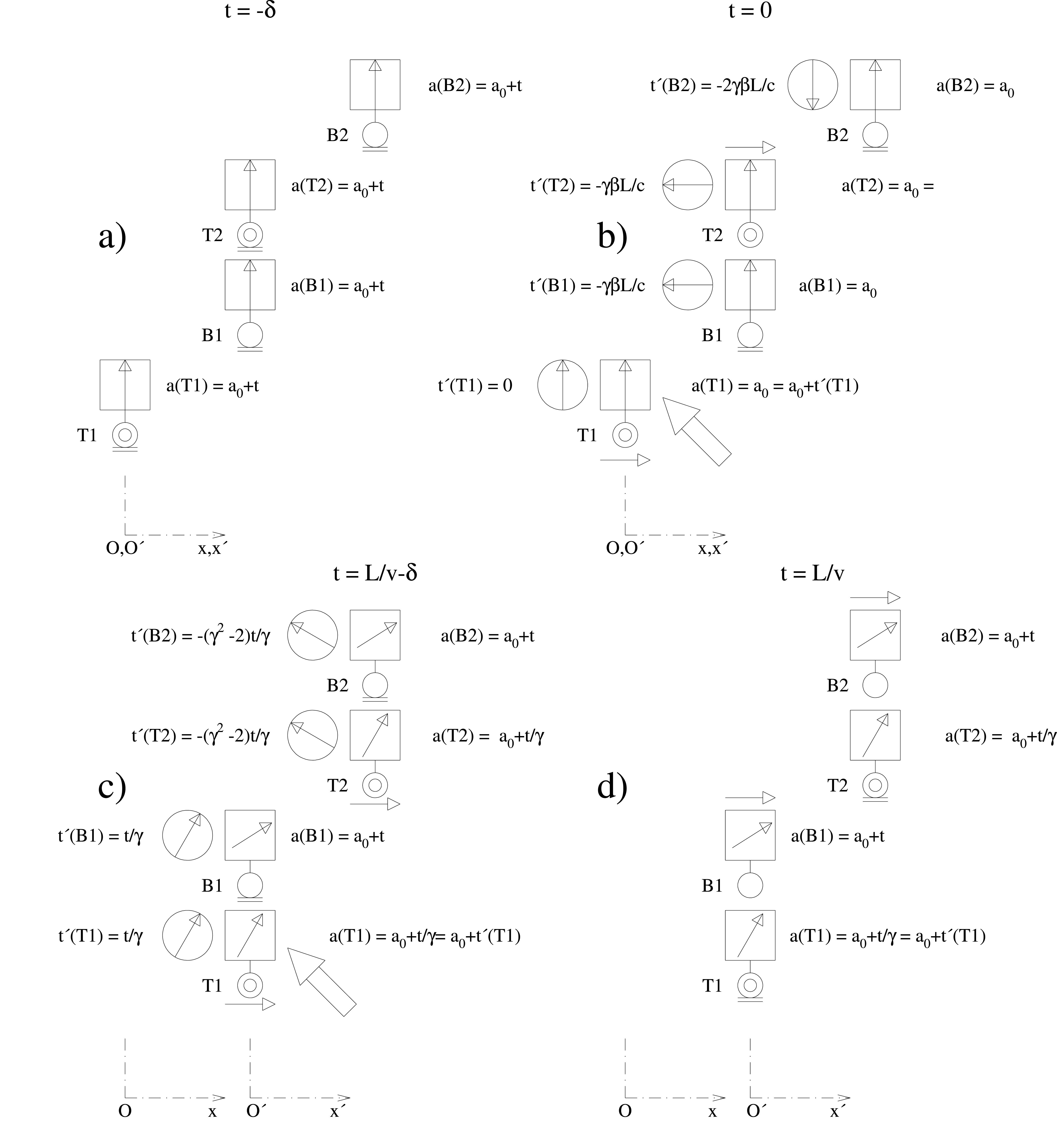}}
\caption{ {\em As Fig. 8 except that B1 is at $x = L$ in S and the origin of S' is aligned
   with  T1. $v = (\sqrt{3}/2)c$, $\gamma = 2$. See text for discussion.}}
\label{fig-fig11}
\end{center}
\end{figure}

\begin{figure}[htbp]
\begin{center}\hspace*{-0.5cm}\mbox{
\epsfysize15.0cm\epsffile{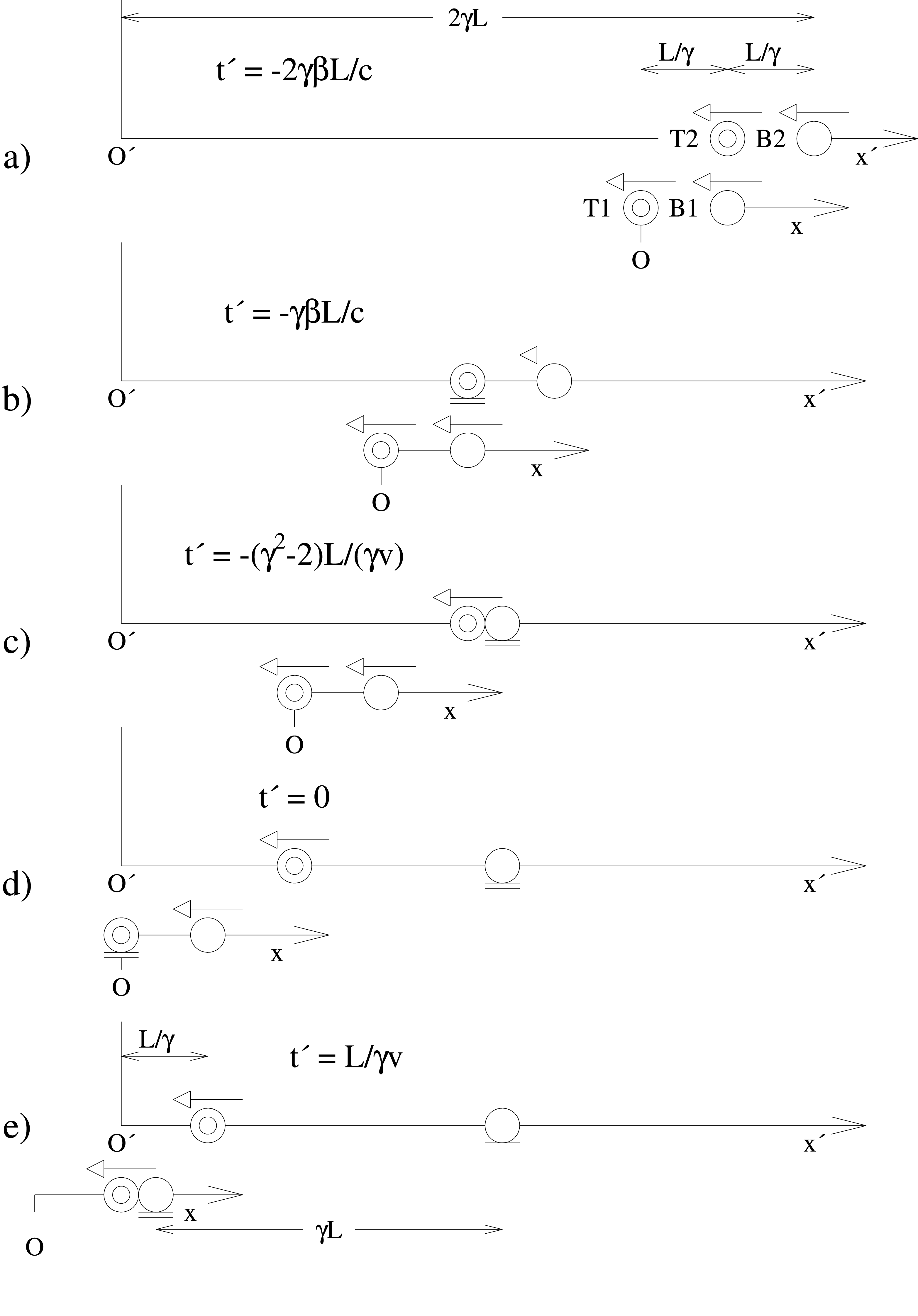}}
\caption{ {\em Event configurations in the travelling frame S' as calculated using (4.1) and (4.2),
 corresponding to the sequence of S (base frame) events shown in Fig. 11. $v = (\sqrt{3}/2)c$, $\gamma = 2$. 
 See text for discussion.}}
\label{fig-fig12}
\end{center}
\end{figure}

    \par The conventional calculation using (4.1) and (4.2) describes correctly the age increments 
    of T1 and T2 while they are at rest in S', but as is evident, in general, from inspection
    of Figs. 9 and 11, and for the special case of T1 by comparing Eqs. (6.1) and (6.2), that
    the generic LT does not always  calculate correctly the ages of the SBs, i.e. the values 
    of $t´ $ calculated may be in manifest disagreement with the ages of the SBs. The exceptions
    to this are are T2 in Fig. 9 and T1 in Fig. 11 ---the SBś that, in each case, are at the origin
    of S'. These correspond to the previously-discussed use of a 'local' LT~\cite{JHFLLT}
     where the coordinate origin in S' lies at the position of the transformed event.

   As previously remarked, the age of T2 is correctly described by (4.1) and (4.2) with the
  choice of spatial coordinate systems shown in Fig. 9  and 10:
   \begin{eqnarray}
    \xr'({\rm T2})& = &\gamma[ \xr({\rm T2}) - v \tr] = 0,  \\
  \tr' & = & \gamma[ \tr - \frac{v \xr({\rm T2})}{c^2}]
    \end{eqnarray}
   where $\tr$ is the time of a synchronised clock at any position in S. These equations 
     may be rewritten in terms of the spatial coordinates of T1 by making use of the identities:
      \begin{eqnarray}
      \xr'({\rm T2}) & \equiv &\xr'({\rm T1}) + [\xr'({\rm T2})-\xr'({\rm T1})] = \xr'({\rm T1}) + L', \\
    \xr({\rm T2}) & \equiv & \xr({\rm T1}) + [\xr({\rm T2})-\xr({\rm T1})] = \xr ({\rm T1}) + L  
  \end{eqnarray}
 to give
    \begin{eqnarray}
    \xr'({\rm T1})+ L & = &\gamma[ \xr({\rm T1})+L - v \tr] = 0,  \\
  \tr' & = & \gamma[ \tr - \frac{v(\xr({\rm T1})+L)}{c^2}]
    \end{eqnarray}   
  where the relation (3.18), $L = L'$ has been used. Setting $\tr = L/v$, (6.9) and (6.10) then give,
    on eliminating $\xr({\rm T1})+L$:
     \begin{equation}
  a({\rm T1})-a_0 = \tr'({\rm T1})  = \frac{\tr({\rm T1})}{\gamma}  = \frac{L}{\gamma v}
    \end{equation}  
    in contrast to (6.1) and in agreement with the TD relation (6.2). Eq. (6.9)
   also gives correctly the age of T1: $ a({\rm T1})- a_0 = \tr'({\rm T1}) = 0$  at $\tr = 0$
     when $\xr({\rm T1}) = -L$. In Table 1 and Fig. 9 it is seen instead that at $\tr = 0$, 
     $ \tr'({\rm T1}) = \gamma \beta L/c  \ne a({\rm T1})- a_0 = 0$
     \par Similarly, with the choice of spatial coordinate systems of Figs. 11  and 12. the age of T2 is
      correctly given by the equations:
    \begin{eqnarray}
          \xr'({\rm T2})- L & = &\gamma[ \xr({\rm T2})-L - v \tr] = 0,  \\
  a({\rm T2}) - a_0 = \tr'({\rm T2}) & = & \gamma[ \tr - \frac{v(\xr({\rm T2})-L)}{c^2}].
    \end{eqnarray}   
   Introducing `local' coordinate systems at the position of T1 in (6.9) and (6.10) or T2 in (6.12)
   and (6.13):
   \begin{eqnarray}
   \tilde{\xr}({\rm T1}) & \equiv & \xr({\rm T1})+L,~~~ \tilde{\xr}'({\rm T1}) \equiv  \xr'({\rm T1})+L, \\
   \tilde{\xr}({\rm T2}) & \equiv & \xr({\rm T2})-L,~~~ \tilde{\xr}'({\rm T2}) \equiv  \xr'({\rm T2})-L
   \end{eqnarray}   
  the form of the generic LT (4.1) and 4.2) is recovered:
   \begin{eqnarray}
         \tilde{\xr}'({\rm T}) & = &\gamma[ \tilde{\xr}({\rm T}) - v \tr] = 0,  \\
     \tr' & = & \gamma[ \tr - \frac{v \tilde{\xr}({\rm T})}{c^2}], \\
          {\rm T} & = &  {\rm T1},~{\rm T1}.  
    \end{eqnarray} 
   \par The simplest manner to express the LT (6.16) and (6.17), which describes a synchronised clock
   at an arbitary position in S' is:
  \begin{eqnarray}
 \tilde{\xr}'({\rm T}) & = & 0, \\
 \tilde{\xr}({\rm T}) & = & v \tr, \\
  \tr & = & \gamma  \tr'. 
    \end{eqnarray} 
  Eqs.(6.19) and (6.20) are equivalent to the space transformation equation (6.16). They
   are the same as in Galilean relativity. The translationally invariant TD relation (6.21)
   obtained by eliminating $\tilde{\xr}({\rm T})$ between (6.16) and (6.17) may be contrasted
   with the correponding equation: $ \tr = \tr'$ of Galilean relativity, that is the limit of
  Eq.(6.21) as $c \rightarrow \infty$. In fact, the full physical content of the the space-time LT
  concerning the transformation of events on the world line of an object at rest in the frame S', is contained
   in Eqs.(6.19)-(6.21). Galilean relativity is recovered on replacing the TD relation (6.21) by the
    absolute time: $\tr_N = \tr = \tr'$ of  Newtonian mechanics. 
   \par It is important to stress the unphysical nature of the S' frame times
    at $\tr = l/v-\delta$ of T1, B1 and B2 in Fig. 9 and of T2, B1 and B2 in Fig. 11 as calculated
    using the generic LT. These clearly do not correspond to the actual ages of the SBs, as
    given by the translationally-invariant TD relation, and shown on the clocks with square
    dials.  
    \par Consider, as another example of such a  mismatch, T1 at $\tr = 0$ in Fig. 9 where
       the generic prediction is $\tr'({\rm T1}) = \gamma \beta L/c$. If this were really 
    the age of the SB, then during the negligibly short collision time of P1 with T1 a fraction
    $1-\exp[\gamma \beta l/(c \tau_D)]$ of the radioactive atoms doping the SB would have
    to decay. In the case of  T2 at $\tr = 0$ in Fig. 11 where  $\tr'({\rm T2}) = -\gamma \beta L/c$
   a fraction  $1+\exp[\gamma \beta l/(c \tau_D)]$ of the atoms would have to `undecay' or equivalently
   this fraction of undecayed atoms would have to be created during the collision of P2
   with T2! Although this unphysical, indeed nonsensical, character of the predictions
   of the generic LT is particularly clear for the case of radioactive clocks, similar 
   considerations apply equally to any type of clock. For, example in the case of
   a spring-driven mechanical clock, the prediction of $\tr'({\rm T2})$ at $\tr = 0$ 
   in Fig. 11, would require the spring to spontaneously rewind during the collision
   of P2 with T2! During more than a century such absurd predictions of generally accepted
   special relativity theory have appeared, without any adverse comment, in text-books and
   the pedagogical literature\footnote{See, for example, Refs\cite{Boughn,Soni}.}

  \SECTION{\bf{The twin paradox and the Minkowski space-time plot}}
    In this section the full outward and return journey of a travelling twin, T,
    is considered. As before, it is assumed that all time intervals during acceleration
     may be neglected as compared to those corresponding to periods of uniform
     motion. Initially, T and the stay-at-home twin R are a rest at the origin
     of the base frame S. T is projected into the frame S' moving with uniform
    speed $v_B$ relative to S. After travelling a distance $L$, T is projected into the
      frame S'' moving with uniform speed $v_B$ in the direction opposite
       to that of S'. On arriving back at the origin of S, T is projected into this frame.
       During T's journey R remains at rest at the origin of S.  
    \par As discussed in Ref.~\cite{JHFMinkP}, the space-time LT is geometrically
     equivalent to the product of an orthogonal projection and a scale
     transformation. The corresponding equations representing T's journey
     are as  follows:
     \par \underline{Outward journey}
     \begin{eqnarray}
       \Delta \bar{X}'_T & = & f(\beta_B)(\Delta X_B \cos \theta -c\Delta T_B \sin \theta), \\
    c\Delta \bar{T}'_T & = & f(\beta_B)(c\Delta T_B \cos \theta -\Delta X_B \sin \theta).
     \end{eqnarray}
 \par \underline{Return journey}
     \begin{eqnarray}
       \Delta \bar{X}''_T & = & f(\beta_B)(\Delta X_B \cos \theta +c\Delta T_B \sin \theta), \\
    c\Delta \bar{T}''_T & = & f(\beta_B)(c\Delta T_B \cos \theta +\Delta X_B \sin \theta).
     \end{eqnarray}
   where $\beta_B \equiv v_B/c$ and
    \begin{equation} 
     \cos \theta \equiv = \frac{1}{\sqrt{1+\beta_B^2}},~~~  \sin \theta \equiv \frac{\beta_B}{\sqrt{1+\beta_B^2}},~~~
       f(\beta_B) \equiv \sqrt{\frac{1+\beta_B^2}{1-\beta_B^2}}.
     \end{equation}
      The world lines of R and T are shown with orthogonal $X_B$ and $cT_B$ axes in Fig. 13 with $\beta_B = 1/3$.
      The two-dimensional rotations in the above equations correspond to orthogonal projections from 
        a point on the world line of T on to the axes of the travelling frame coordinates. Note that 
        for the transformations (7.1) and (7.2), the $c\bar{T}'_T$ axis is obtained from the $T_B$ axis 
       by an anti-clockwise rotation through the angle $\theta$, and the $\bar{X}'_T$ axis from the
         $X_B$ one by clockwise rotation through the angle $\theta$. As pointed out in Ref.~\cite{JHFMinkP},
         in Minkowski's original paper a sign error was made in plotting the directions of the travelling frame
          axes so that, for example, the $T'_T$ axis is superimposed on the corresponding world line.
          This error has persisted since in text books and the pedagogical literature, including, as will be
         seen below, 
         many discussions of the twin paradox. As shown in Fig.13, the combination of orthogonal
         projection and scale transformation is equivalent to projecting at the angle $\phi_t$ onto
         the $c\bar{T}'_T$ axis where~\cite{JHFMinkP}:
        \begin{equation}
         \tan \phi_t = \frac{2 \beta_B}{\sqrt{1-\beta_B^2}(\sqrt{1+\beta_B^2} - \sqrt{1-\beta_B^2})}.
        \end{equation}      
         \par The world lines of T and R in the rest frame of T (S' on the outward journey, S'' on the
           return one) are shown on orthogonal $c\bar{T}'_T$ versus  $\bar{X}'_T$ and
         $c\bar{T}''_T$ versus  $\bar{X}''_T$ plots in Fig. 14, also for  $\beta_B = 1/3$.
          The world lines of T are inclined at an angle $\theta' = {\rm arctan} \gamma_B \beta_B$ to the
        $c\bar{T}'_T$ or  $c\bar{T}''_T$ axes.
          \par The DAE is read off from the  different time intervals shown in Fig. 13. Denoting the age
        increments of T and R by $\Delta a_T$ and $\Delta a_R$ respectively then
          \begin{eqnarray}
           \Delta a_T & = & \Delta \bar{T}'_T +\Delta \bar{T}''_T 
            = 2\Delta \bar{T}'_T =f(\beta_B)\frac{\cos 2\theta}{cos \theta} \Delta T_B \nonumber \\
             & = & \sqrt{\frac{1+\beta_B^2}{1-\beta_B^2}}\sqrt{1+\beta_B^2}
               \frac{1-\beta_B^2}{1+\beta_B^2} \Delta T_B
           = \frac{\Delta T_B}{\gamma_B} =\frac{\Delta a_R}{\gamma_B}. 
           \end{eqnarray}  
 
\begin{figure}[htbp]
\begin{center}\hspace*{-0.5cm}\mbox{
\epsfysize15.0cm\epsffile{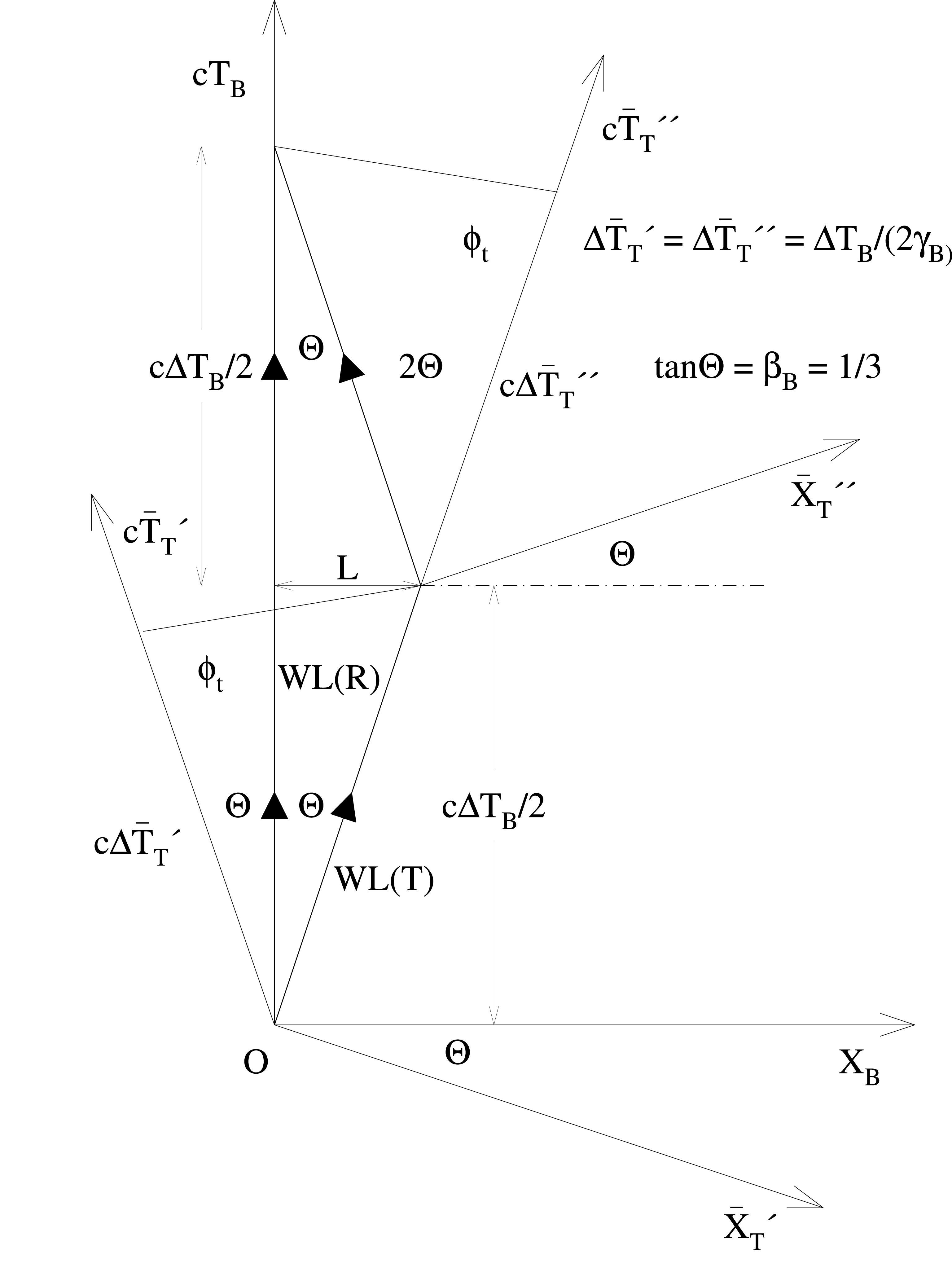}}
\caption{ {\em World lines of R (WL(R)) and T (WL(T)) in the frame S. Coordinate axes of the
  frames S' and S'' as predicted by the LT (7.1)-(7.4) are also show.  See text for discussion.}}
\label{fig-fig13}
\end{center}
\end{figure}  

\begin{figure}[htbp]
\begin{center}\hspace*{-0.5cm}\mbox{
\epsfysize15.0cm\epsffile{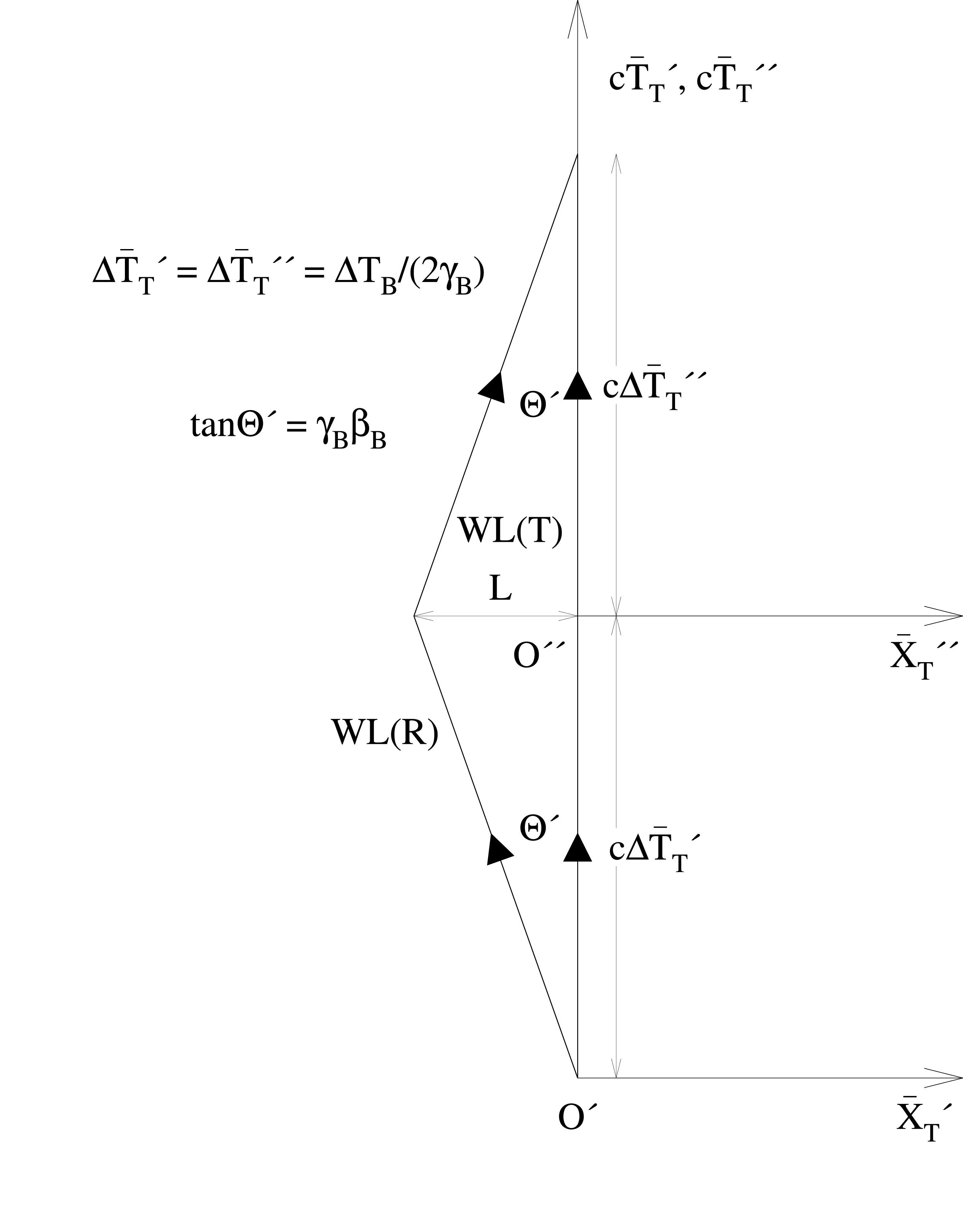}}
\caption{ {\em World lines of R (WL(R)) and T (WL(T)) in the frames S' and S'' as predicted by the LT (7.1)-(7.4).
 See text for discussion.}}
\label{fig-fig14}
\end{center}
\end{figure} 

       \par The space-time description of Figs. 13 and 14, in the base and travelling frames
      respectively, is now compared with the standard `resolution'of the twin paradox as presented
       in text books and the pedagogical literature. This can be found for example in Taylor and Wheeler's
      book `Spacetime Physics' ~\cite{TW} or in Chapter 3 of Marder's book on the twin paradox~\cite{Marder}.
       As before, time intervals in which the travelling twin is accelerated or deceletrated
      are neglected as compared to time intervals corresponding to uniform motion. It will now 
      also be assumed that T remains at rest for a short period after completing the outward journey
       and before starting the return one. The space-time plot on which the discussion is based is that
        shown in Fig.15 (c.f. Fig.71 of Ref~\cite{TW} or Fig. 17 of Ref~\cite{Marder}). Usually the vertical
        axis is labelled as $c\tr$ where $\tr$ is the age increment of the stay-at-home twin R, however detailed 
        geometrical analysis of the plot (not performed in Refs~\cite{TW,Marder}) shows that this
        is not possible, so that the symbol $\tilde{\tr}$ is used in Fig. 15 to label the vertical axis.
        The relation connecting $\tr$ to  $\tilde{\tr}$ is found below. Notice that the error in the
        original Minkowski paper~\cite{Minkowski} of drawing the $c \tr'$ axis (where $\tr_T = \tr'$ is T's age
       increment during the outward journey) along the world line of T, and the $c \tr_T = c \tr''$ axis
        along the world line of T, during the return journey, is made.  Compare with Fig. 13 where the 
        correct directions of these axes as predicted by the LT (7.1) and (7.2) or (7.3) and (7.4)
        are shown. The $\xr'$ and $\xr''$ axes are also incorrectly drawn (they are reflected in the
         $\xr$ axis as compared to the correct directions shown in Fig. 13). In order to `resolve' 
         the paradox it is now assumed the the age increment of R, as  observed from T, is subject
         to the TD effect for a moving clock in R's frame as compared
         to a clock at rest in T's frame.
          If $\Delta \tr_T $ is the total age increment for T while in uniform motion
         relative to R, then the segment OU in Fig. 15 represents $c\Delta \tr_T/2 $. The next 
         assumption is that the line PU parallel to the $\xr'$ axis is a `line of simultaneity'
         in the frame S' and that OP represents an age increment in the frame S given by the TD effect
          as $\Delta \tr_1 = \Delta \tr_T/(2 \gamma)$.
          Since the geometry of Fig. 15 gives:
             \begin{equation}
            \frac{{\rm OU}}{{ \rm OP}} = \frac{\sqrt{1+\beta^2}}{1-\beta^2}
        \end{equation}
    it follows that the $c\tilde{\tr}$ segment OP must be multiplied by the factor
     $F \equiv \sqrt{1+\beta^2}/\sqrt{1-\beta^2}$ in order the represent correctly the age increment
        $ \Delta \tr_1$. This means that in Fig. 15:
       \begin{equation}
        \tr =  \sqrt{\frac{1+\beta^2}{1-\beta^2}}\tilde{\tr} \equiv  F \tilde{\tr}. 
   \end{equation}
    On the return journey the `line of simultaneity' is SV so that ST also represents the age increment 
    $\Delta \tr_T/(2 \gamma)$. According to the geometry of Fig. 15, the total age increment for R, $\Delta \tr$,
    as seen by T during her two intervals of uniform motion, is then given by:
      \begin{eqnarray}
  c\Delta \tr & = & F c\Delta \tilde{\tr} = 
   F({\rm OQ}+{\rm RT}) \nonumber \\
    & = & 2 F ({\rm OU} \cos \theta) = 
 2 \sqrt{\frac{1+\beta^2}{1-\beta^2}}\frac{c \Delta \tr_T}{2} \frac{1}{\sqrt{1+\beta^2}}  \nonumber \\
  & = & \gamma c\Delta \tr_T
 \end{eqnarray}
  which corresponds to the TD effect for T in motion as observed from R at rest. Since this agrees with R's
   calculation of the DAE it is presented as the `solution' of the paradox. However, this implies that 
    the age of R (as observed by T) increases by the time interval $\Delta \tr_2$ where:
   \begin{equation}
      c\Delta \tr_2 = F c\Delta \tilde{\tr} = F({\rm PQ}) = \sqrt{\frac{1+\beta^2}{1-\beta^2}}
        \frac{c \Delta \tr}{2} \sin \theta \tan \theta = \frac{\gamma \beta^2 c \Delta  \tr_T}{2}
   \end{equation}
    during T's negligibly short period of decelaration before coming to rest in
     S at the end of the outward journey. Similarly, the age of R (as observed by T)
     increases by the same amount, corresponding to the $c \tilde{\tr}$ interval RS in Fig. 15
     during the negligibly short period during which T is accelerated from the point V at the
     beginning of her return journey. 
       \par A similar interpretation was obtained by Muller~\cite{Muller} by direct use of the
    generic inverse LT (4.3) and (4.4). Substituting $\xr' = -v \tr'$ (corresponding to $\xr = 0$)
    in (4.4) gives:
     \begin{equation}
  \tr_{{\rm O}} = \gamma(\tr_T+\frac{v \xr'}{c^2}) = \frac{\tr_T}{\gamma}
  \end{equation}
  where $\tr_T$ is the traveller's proper time (here  $\tr_T = \tr'$) and $ \tr_{{ \rm O}}$
   is the age increment of R at the end of the outward journey. Neglecting the time interval
    corresponding to UV in Fig. 15, Muller assumes that the relation between $\tr_T$ and $\tr$
    at the beginning of the return journey is given by the first member of (7.12) with the replacement $v \rightarrow -v$:
   \begin{equation}
  \tr_{{\rm R}} = \gamma(\tr_T-\frac{v \xr'}{c^2})
  \end{equation}
    where now  $\tr_T = \tr''$ and $\tr_{{ \rm R}}$
       is interpreted as the age of R (as observed by T) at the beginning
   of the return journey. Combining (7.12) and (7.13) gives:
     \begin{equation}
   \tr_{{\rm R}} -  \tr_{{\rm O}} = 2 \Delta t_2 = 2 \gamma \beta^2 \tr_T = \gamma \beta^2 \Delta \tr_T
    \end{equation}
    in agreement with (7.11) above.     
    \par If $L$ and $L'$ are the separations of R and T, in the frames
      S and S' respectively, at the end of the outward journey just before
    T's deceleration phase, use of the relations:
       \[ -\xr' = L' = \frac{v \Delta \tr_T}{2},~~~ L = \frac{v \Delta \tr}{2} \]
       to eliminate $\Delta \tr_T$ and $\Delta \tr$ from the TD relation (7.10) gives the 
       `length contraction' relation:
      \begin{equation}
          L' = \frac{L}{\gamma}.                                                                                                                                  
       \end{equation}
     As discussed in the previous section, the DAE of R and T is then explained as the result 
    of the smaller distance covered by T on the assumption (contrary to the reciprocity relation
      (3.21)) that the magnitude of the velocity of R relative to T in the frame S' is the same
       as the initially fixed velocity, $v$, of T relative to R in the frame S.
       \par  The `standard
       solution' of the twin paradox is summarised in Fig. 15, previously discussed and in
       Figs. 16, 17 and 18.  In Fig. 16 are shown the world lines of R as viewed from the proper
      frame of T, which is successively S,S' S, S'' and S. The remarkable feature of this plot
      is the change of the separation of T and R from $L/\gamma$ to $L$ during the deceleration
       of T at the end of the outward journey and from $L$ to $L/\gamma$ during the acceleration
        phase at the begining of the return journey.  For short acceleration and deceleration
        phases the displacement of R relative to T occurs with a superluminal velocity, that
       tends to infinity as the ratio of the periods of acceleration or deceleration
       to the periods of uniform motion tends to zero. 

\begin{figure}[htbp]
\begin{center}\hspace*{-0.5cm}\mbox{
\epsfysize15.0cm\epsffile{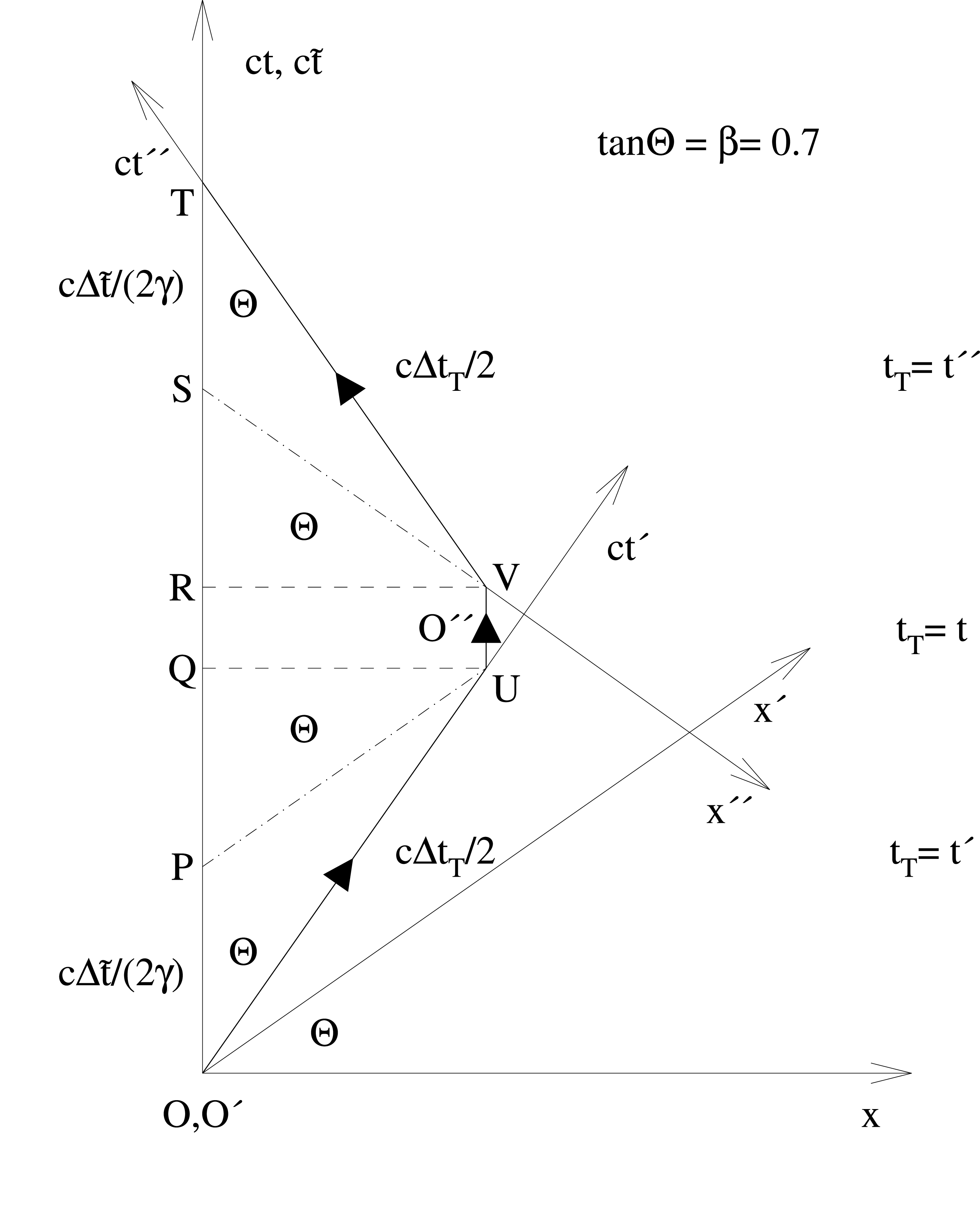}}
\caption{ {\em Space-time plot used in the conventional
 discussion of the twin paradox based on the generic
LT (4.1)-(4.2). Note that the $t'$ and $t''$ axes are incorrectly drawn along the world
 lines. (Compare with Fig. 14). See text for discussion}}
\label{fig-fig15}
\end{center}
\end{figure}
 
\begin{figure}[htbp]
\begin{center}\hspace*{-0.5cm}\mbox{
\epsfysize12.0cm\epsffile{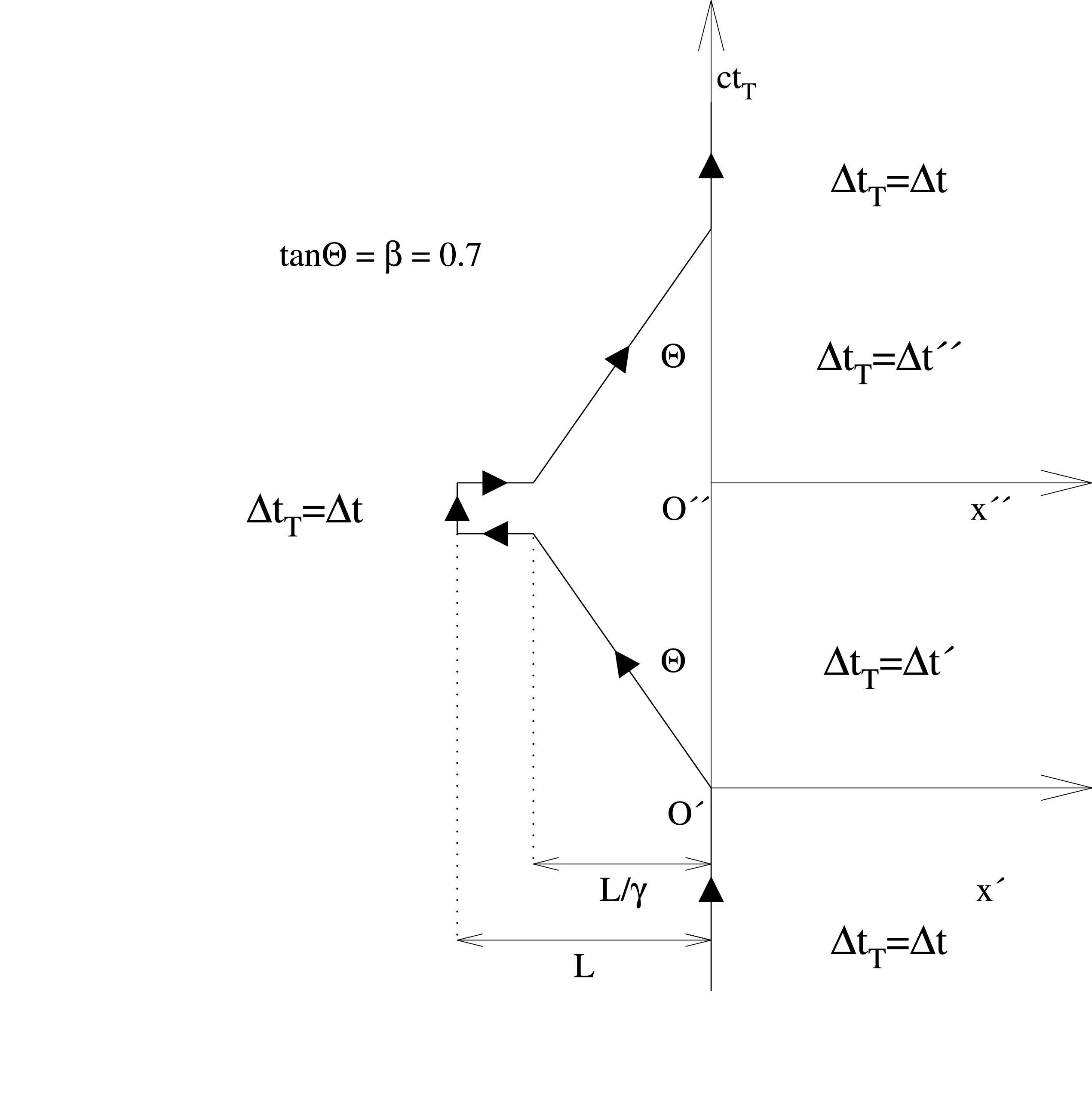}}
\caption{ {\em Space time plot of R in the traveller's rest frame in the 
 conventional discussion of the twin paradox as in Ref.~\cite{McRea}. See text for discussion}}
\label{fig-fig16}
\end{center}
\end{figure}

\begin{figure}[htbp]
\begin{center}\hspace*{-0.5cm}\mbox{
\epsfysize12.0cm\epsffile{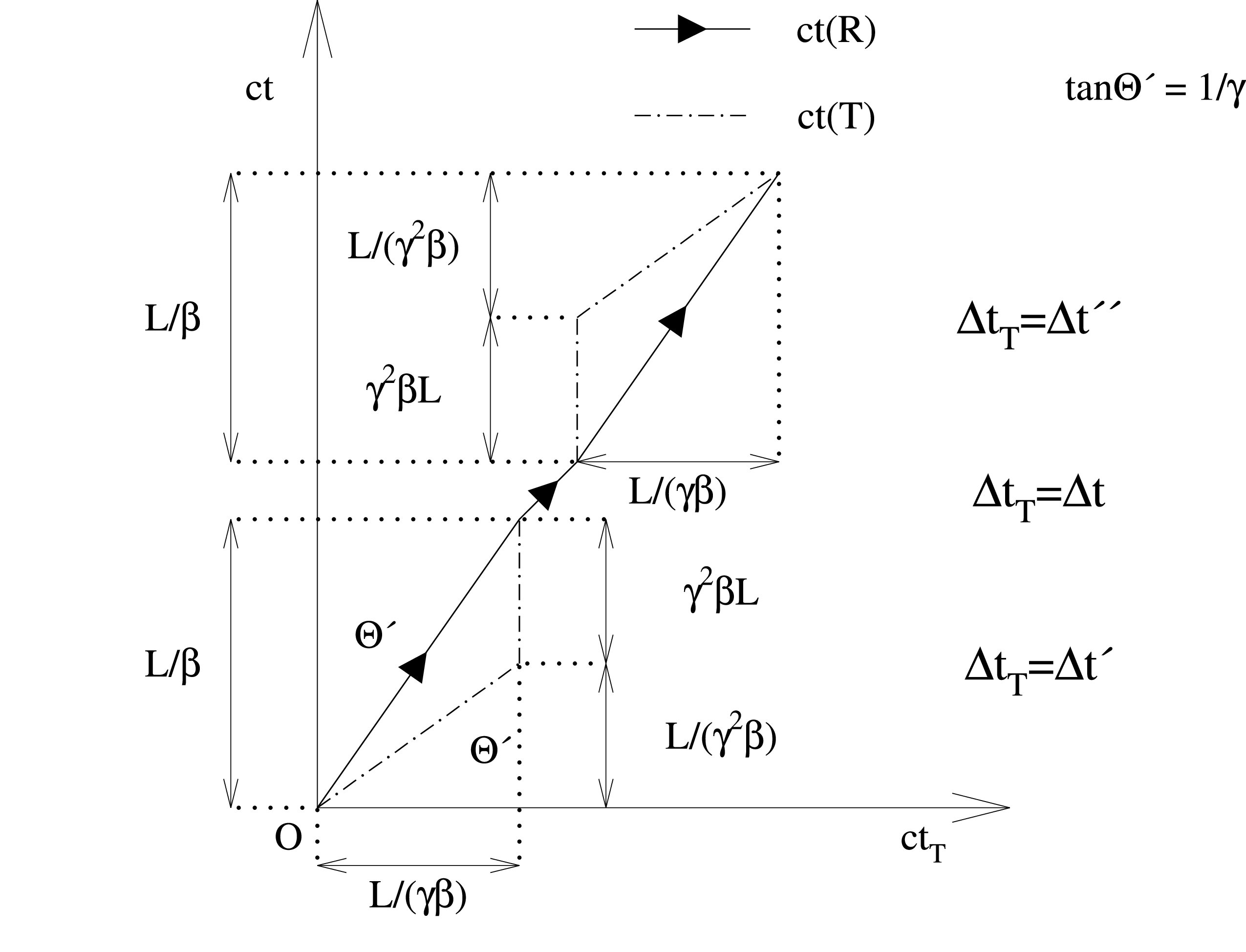}}
\caption{ {\em Plots of times in the frame S as observed by R: $t(R)$, and by T: $t(T)$,
  as a function of $t_T$ according to the calculation of Ref.~\cite{Muller}. See text for discussion.}}
\label{fig-fig17}
\end{center}
\end{figure}
 
\begin{figure}[htbp]
\begin{center}\hspace*{-0.5cm}\mbox{
\epsfysize15.0cm\epsffile{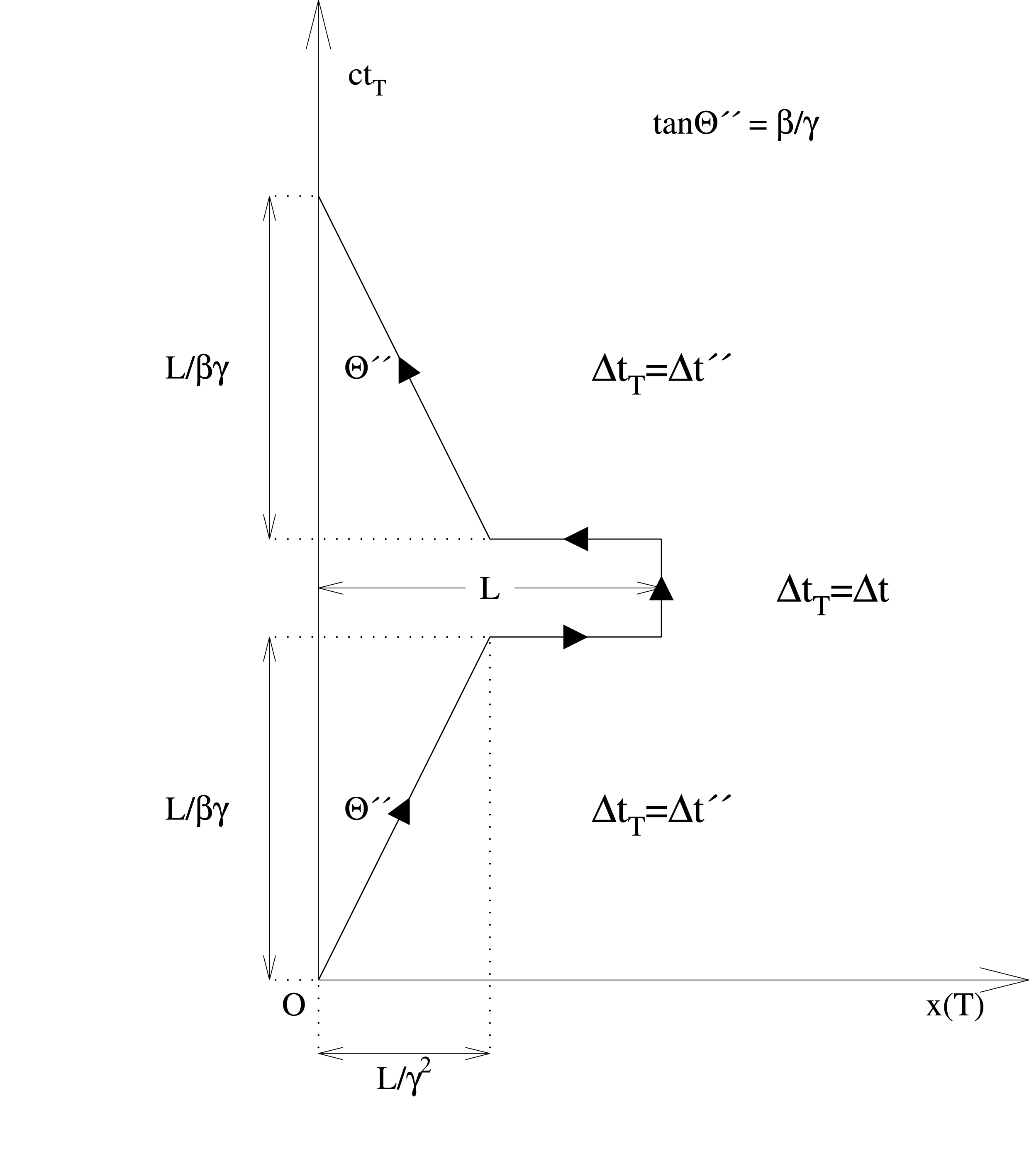}}
\caption{ {\em Plot of $ct_{{\rm T}}$ as a function the separation of R from T in the frame S, as observed by T: $x(T)$,
 according to the calculation of Ref.~\cite{Muller}. See text for discussion.}}
\label{fig-fig18}
\end{center}
\end{figure}

       \par In Fig. 17, taken from Ref~\cite{Muller} is shown a plot of $c\tr$ versus  $c\tr_T$
        where $\tr(R)$ (solid line) is R's age as observed by R and  $\tr(T)$ (dashed line)
        is  R's age as observed by T, according to the calculation in Eqns (7.12)-(7.14).
        During the infinitesimally short deceleration period of T at the end of the outward
        journey R ages (according to T) by an amount $\gamma^2 \beta L/c$. R also ages by
        the same amount (according to T) during the acceleration period at the beginning
        of the return journey.
        \par In Fig. 18 is shown the $\xr$-coordinate of T, $\xr(T)$, as a function of
        $\tr_T$ which corresponds to the dot-dashed line of Fig. 17. Near the end of the
        outward journey, while T is still in motion, the separation of R and T is
        $L/\gamma^2$. During the short deceleration period the separation increases to $L$.
        During acceleration at the beginning of the return journey the separation shrinks
        again to $L/\gamma^2$. The angle $\theta''$ in Fig. 18 is given by $\tan \theta'' = \beta/\gamma$
        corresponding to a slowed-down relative velocity of R and T, as seen by T, of
        $v(T) = \xr(T)/\tr_T = c \tan \theta'' = v/\gamma$.
        \par The scenario shown in Fig. 16 has been graphically described by McRea~\cite{McRea}
         (the travelling twin is here denoted by M):
        \newline
        \par `At the event of M completing the part of his outward journey, performed with speed $V$,
         observer R plots M as at a distance $X(=VT)$ from himself. (We say ``plot'' instead of
         ``see'' to avoid complications about times of travel of light in the process of seeing)
         Owing to the Lorentz contraction, M at the same event plots R as at distance $\alpha X$
         from himself where $\alpha = \sqrt{1-(V/c)^2}$. During the interval, $t$, when R,M are both
         at rest in F, they must plot each other as at distance X. But as soon as M has again 
        acquired uniform speed $V$, this time towards R, he again plots R as at distance $\alpha X$.
         \par R's account of M's journey relative to himself is, still neglecting quantities small
         compared with $X$ and $T$: M moves out and back through distance $X$ at speed $V$;
         total time $2T$. On the other hand, M's account of R's journey relative to himself is:
         R moves out the distance $\alpha X$ at speed $V$ then R ``cheats'' by moving out to 
        the distance $X$ and back again to the distance $\alpha X$ in negligible time and 
         finally finally completes his return at speed $V$; total time $2 \alpha T$. This
         is less that $2T$ and is the usual result.'
      \newline
       \par This standard solution of the paradox will now be examined for mathematical and
       logical consistency, before analysing it according to the nomenclature and notation
       for space and time coordinates introduced in Section 3 above.
        \par Qualitative arguments based on the incorrectly drawn Minkowski plot of Fig. 15 are 
    to be found in many places in the the literature~\cite{TW,Marder,Lowry}, but nowhere, to the 
        present writer's best knowledge, is the geometrical problem worked out in full detail
        \footnote{For example, In Ref.~\cite{TW} the solution is presented according to Eqns. (7.12)-(7.14)
     above as in Ref.~\cite{Muller} without any reference to the detailed geometry of the Minkowski
      plot.} If it had been, the geometrical inconsistency of assuming that
      ${\rm OP}/{\rm OU} = 1/\gamma$, according to TD, when the vertical axis represents $ct$ would
      have been noticed. If the vertical axis is used to represent `` time in S as observed from S'~''
      as in the derivation of Eqn.(7.10), it cannot also represent `` time in S as observed from S~''
      as is usually assumed in drawing the Minkowski plot. Indeed identification of the line segment 
      OU with $c \Delta \tr_T/2$ instead of $\sqrt{(c \Delta \tr/2)^2+L^2}$ means the interpretation
      of the plot must be changed. In fact, the angle $\theta$ is now completely arbitary. If it
      is assumed that ${\rm OP}/{\rm OU} = 1/\gamma$ to correctly describe the TD effect, any 
       change in $\theta$ will be automatically compensated by the scale factor $F$ that must
       be applied to the time coordinate, since the geometry of Fig. 15 does not correctly
       predict the TD effect if $\tan \theta = \beta$. 
       \par As in the analogous case discussed in the previous section, the sudden `age increases'
        of R as seen by T, during deceleration and acceleration, and associated with PQ and RS in Fig. 15
        are spurious, resulting from incorrect use of the space-time LT. If there is a local
        clock at rest in the frame S, synchronised with R's clock, at the position U in Fig. 15,
        at the end of the outward journey, this clock will be in advance of T's clock by
        an amount $L(1-1/\gamma)/v$, and both T and local observer in S at U will agree that this
        is the case. If the periods of deceleration at the end of the outward journey and
        acceleration at the beginning of the return journey are very short, T's clock will
        still be retarded by essentially the same amount compared to the local clock in S at the
        start of the phase of uniform motion at the beginning of the return journey when T 
       is in the frame S''. Since the times recorded by the local clock at U and R's 
       clock are identical there is, contrary to the standard solution, no change of R's age
       `as seen by T' during the deceleration and acceleration phase. In order to correctly
        describe the TD effect during the return journey, additive constants must be included
        in the generic LT (4.1) and (4.2) in order to correctly represent the intial conditions
        of the return journey as just described. Thus the generic LT must be modified by including 
        time offsets $t_0$ and $t''_0$ \footnote{As this is no longer the generic LT of conventional
         text-book special relativity the coordinates no longer use a roman font.} in order to
         correctly describe the transformation between the frames S and S'':
        \begin{eqnarray}
        x'' & = & \gamma[x+v(t-t_0)], \\
        t''-t''_0 & = & \gamma[t-t_0+\frac{v x}{c^2}].
         \end{eqnarray}
         The time offsets must be chosen so that $t''= L/(\gamma v)$ and  $t= L/ v$ when $x''=0$ and
         $x =  L$. This requires that:
      \begin{eqnarray}
        t_0 & = &\frac{2L}{v}, \\
        t''_0 & = &\frac{2L}{\gamma v}
    \end{eqnarray}
     so that (7.16) and (7.17) may be written as:
        \begin{eqnarray}
        x'' & = & \gamma[x-2L+vt], \\
        t'' & = & \gamma[t+\frac{v(x-2L)}{c^2}].
         \end{eqnarray}
     Eliminating $x-2L$ between (7.20) and (7.21) gives:
    \begin{equation}
      t_T = t'' = \frac{t}{\gamma}
    \end{equation}
    which describes the TD effect during the return journey. Since the local clock at U is 
    synchronised with the local clock at R and therefore measures R's age, there are only two times
    in the problem: $t$, which is the proper time of R, equal to his age increment during
     T's journey, and $t_T$ the proper time of T,  equal to his age increment during his
     journey, which is equal to $t'$ on the outward and equal to $t''$ on the return journey. As is
     also clear from inspection of the correct space time plots in Figs. 13 and 14, there
     are no sudden changes of ages and spatial separations as in the standard solution
     shown in Figs. 15-18. In particular there is no distinction between `the age of R'
     and `the age of R as observed by T'. The latter quantity is a spurious mathematical
     artifact, which, as will now be seen, results form incorrectly identifying space-time
     coordinates in the physically distinct reciprocal experiment with transformed coordinates
     of the primary experiment.
      \par The standard solution is now considered in the light of the classification  of space-time
        experiments introduced in Section 3 above. In the twin paradox problem, S, the rest frame
        of R, is the base and source frame and S' or S'' is the travelling and target frame. S is 
        the subject frame considering R as an observer and S' or S'' are subject frames
         considering T as an observer. The initial conditions of the problem ($L$ and $v$) are therefore
        specified in the frame S. A single LT (or its inverse) describes the problem on the outward
       journey, another one on the return journey. On the outward journey it is the LT
       for which the spatial coordinates of T are $x' =0$ in S' and $x = vt$ in S, and $t_T = t'$,
         whereas on the return jorney it is one for which the spatial coordinates 
         of T are $x'' =0$ in S'' and $x =2L-vt$ in S, and $t_T = t''$.
        The quantity $\tr(T)$, `the time in S as observed from T' corresponding to the dot-dashed line in 
         Fig. 17, does not occur in the description just given. How then, is it introduced in
        the standard solution? It is done by introducing the LT with $\xr = 0$ 
        and $\xr'= -v \tr'$ into the problem. In the language introduced in Section 4, this is the
         LT appropriate to the experiment which is reciprocal to the one specified above. In 
        the reciprocal experiment S' is the base and source frame in which the initial
       conditions are specified, and S is the travelling and target frame. The LT is 
       therefore one which applies to an experiment which is physically quite
       distinct, having nothing to do, as is assumed in the standard solution, with
       observation of R by T in the primary experiment.
       Indeed, if S is the base frame and S' the travelling frame
       the equation of motion of R in S' is given by Eqn(3.21) as 
        $x' = -\bar{v}´_T t' = -\gamma v t'$ not  $\xr'= -v \tr'$ as in Eqn(7.12) above.
        \par Considering the outward journey, the following space-time LTs (and their inverses)
         describe the primary experiment, in which S is the base and source frame
         and S', the travelling frame and object frame, 
       and the reciprocal experiment in which S' is the base and source frame and S is the 
        travelling and object frame\footnote{Uncapitalised symbols are used throughout for
         simplicity since the equations are invariant under replacements such 
         as $x \rightarrow X$, $x' \rightarrow X'$ etc corresponding to the introduction of
         observers in the frames S, S' respectively.}:
          \par \underline{Primary experiment}
           \par Transformation
       \begin{eqnarray}
       \x'(\Tr) & = & \gamma[x(\Tr)-vt(\Rr)] = 0~~\rightarrow x(\Tr) = v t(\Rr), \\
        \tb'(\Tr) & = & \gamma[t(\Rr)-\frac{v x(\Tr)}{c^2}]~~\rightarrow  \tb'(\Tr) = \frac{t(\Rr)}{\gamma}.
       \end{eqnarray}
         \par Inverse Transformation 
           \begin{eqnarray}
       x(\Tr) & = & \gamma[\x'(\Tr)+v \tb'(\Tr)] = \gamma v \tb'(\Tr)=  v t(\Rr),  \\
        t(\Rr) & = & \gamma[\tb'(\Tr)+\frac{v \x'(\Tr)}{c^2}] = \gamma\tb'(\Tr).
       \end{eqnarray}
         \par \underline{Reciprocal experiment}
           \par Transformation
       \begin{eqnarray}
       \x(\Rr) & = & \gamma[x'(\Rr)+vt'(\Tr)] = 0~~\rightarrow x'(\Rr) = -v t'(\Tr), \\
        \tb(\Rr) & = & \gamma[t'(\Tr)+\frac{v x'(\Rr)}{c^2}]~~\rightarrow  \tb(\Rr) = \frac{t'(\Tr)}{\gamma}.
       \end{eqnarray}
         \par Inverse Transformation 
           \begin{eqnarray}
       x'(\Rr) & = & \gamma[\x(\Rr)-v \tb(\Rr)] = -\gamma v \tb(\Rr)=  - v t'(\Tr),  \\
        t'(\Tr) & = & \gamma[\tb(\Rr)-\frac{v \x(\Rr)}{c^2}] = \gamma\tb(\Rr).
       \end{eqnarray}
       The transformation equations for the reciprocal experiment are obtained from those of
      primary one by the exchanging primed and unprimed quantities and the labels R and T, and by
     making the replacement $ v \rightarrow -v$. It is important to note that identical predictions
     are obtained by use of a transformation and its inverse, and, as previously pointed out in Section 4,
     these predictions do not depend on the choice of source and target frames (i.e. they
     are invariant with respect to exchange of barred and unbarred quantities). Notice that the spatial
     coordinates of the base frame object, R in S in the primary experiment, and T in S' in the
     reciprocal experiment, do not occur in the corresponding transformation equations.
      \par The error made in the standard solution is now quite clear. The LT (7.27) and (7.28) 
       of the reciprocal experiment is mistakenly identified with the inverse transformation
      (which is actually (7.25) and (7.26))
     of the primary experiment. Notice that disregarding completely the labels and subscripts in
     the coordinates, as in the generic inverse LT (4.3) and (4.4), these transformation
      equations become identical.  The quantities $\tr(T)$ and $\xr(T)$ plotted in Figs. 17 and 18 
     respectively are therefore mathematically meaningless and have no physical significance.
      The spurious LC effect of Fig. 16 results, as explained, in Section 6 above above, from incorrect use of
       the generic LT (4.1) and (4.2), that correctly describes a synchronised clock in S' only 
      at $\xr ' = 0$, to describe a synchronised clock for which $\xr ' \ne 0$. The correct world
      lines of R and T and S and S' or S'' are shown in Figs. 13 and 14 respectively. No age changes
       or sudden spatial displacements occur during the short periods of acceleration or
       deceleration.
   \SECTION{\bf{ Summary and closing remarks}}
     The twin paradox of special relativity is defined in the Introduction.  
     In the `prologue' of Section 2 some different methods to measure the length
     of a train are considered: either statically or by using in various ways
     the formula $\Delta x = v \Delta t$ relating constant velocity to space and time intervals.
     Of particular importance for the correct resolution of the twin paradox is the case 
     where observers in two different inertial frames, with clocks running at different
     rates, compare their estimations of the length of the train. 
    \par In Section 3 a general nomenclature and notation for an unambigous description of different
     space-time experiments is introduced. The most important novel concepts are those of a
     `base frame' and a `travelling frame'. The former is an inertial frame (say S) relative to
      which another inertial frame (say S') is measured (or defined) to move with speed $v_B$.
      In this case S' is the `travelling frame'. Important related concepts are those of a 
       `primary experiment', another experiment with a `reciprocal configuration', and a 
       `reciprocal experiment'. In the above example the primary experiment is that in 
        which S is the base frame and S' the travelling frame. Assuming that S' moves along
       the $x,x'$-axis in S, then in an experiment with a reciprocal configuration, S' is the
      base frame and S is a travelling frame moving with speed $v'_B$ along the 
     negative  $x,x'$-axis in S'. If $v_B = v'_B$ the latter experiment is said to be
     the reciprocal of the former and {\it vice versa}. Crucial for the correct
     interpretation of the twin paradox is that the primary experiment and its reciprocal
     are {\it physically independent} ---in particular the base frame space-time coordinates in
     the primary experiment and its reciprocal are not connected by the space-time LT.
      It is the false
     assumption that this is the case which underlies the spurious `standard resolution' of the
     paradox discussed in Section 6.
     However, base frame velocity, or energy-momentum configurations in an experiment and its
       reciprocal are connected by the kinematical (energy-momentum) LT or the base frame
       velocity addition formula (3.28).
      \par Other less important attributes of inertial frames introduced in Section 3 are
        `subject' and `object' frames specifying in which frames observations or measurements
       are made or predicted and `source' or `target' frames i.e. whether a LT or 
        an inverse LT is used to relate space and time coordinates in different inertial
        frames. All physical predictions are independent of whether a frame is classified
       as  `subject' or `object' or by exchange of `source' and `target' frames.
        The use of this nomenclature is illustrated in Section 3 by working out several
        examples ---Time dilation, Lorentz invariance of length intervals, velocity
         addition formulas, transverse and longitudinal photon clocks, Einstein's 
        1905 discussion~\cite{Ein1} of `relativity of simultaneity' and Einstein's
         train-embankment thought experiment~\cite{EinTETE}.
        \par In Section 4 the important distinction between inverse transformations 
          and reciprocal experiments is discussed in detail. Also discussed in Section 4
         is the related `Measurement Reciprocity Postulate' that provides a simple
        axiomatic derivation of the LT~\cite{JHFSTP1,JHFHPA} without any consideration
       of light signals or electromagnetism.
         \par The twin paradox is discussed at length in Section 5 by analysing three 
          different, but related, thought experiments in each of which the periods of
          acceleration or deceleration are negligibly short in comparison with
          periods of uniform motion. In these analyses the nomenclature and notation
         introduced in Section 3 is systematically applied. In the first experiment,
         the stay-at-home twin is identified with a charged pion, created in a high
         energy interaction, that is slow and comes rapidly to rest in the production
         target. The travelling twin is identified with another charged pion, created
         in the same event, but with high energy, that travels a distance of 100m before
         colliding head-on elastically with another charged pion at rest, which projects
       it into the same rest-frame as its stay-at-home sibling. In this example the
       decay lifetimes of the pions measure their age and make manifest the
       differential aging effect ---the age of the travelling pion when it is
       projected into the lab frame is much less than that of its stay-at-home sibling.
       The second experiment is similar to the first except that the pions are replaced by
        macroscopic objects: `Space Billiards' doped with a radioactive substance serving
        as an internal clock to measure their age. They are projected into and out of
        the travelling frame S' by head-on elastic collisions with identical space billiards.
        The analysis of the experiment shows that the physical phenomenon underlying the
        experiment is the relative velocity transformation formula derived in Section 3
         ---if the speed of the space billiard is $v_B$ in the base frame S, it is
          $\gamma_B v_B$ in the travelling frame S'. The third experiment is a variation
          of the first in which a travelling $\pi^-$ induces the process: $\pi^-p \rightarrow \pi^0 n$. 
         The comparison of this primary experiment with its reciprocal, in which the $\pi^-$ remains
         a rest and the proton is in motion, makes clear the 
        physical independence of the two experiments.
        \par In Section 6 the `Space Billiard' experiment of Section 5 is analysed according to 
           the standard solution of the twin paradox where the DAE is found to be a consequence
            of a spurious `length contraction' effect resulting from misinterpretation of
         the LT as explained in Refs~\cite{JHFLLT,JHFUMC,JHFCRCS,JHFACOORD} and recalled in
         Section 6 of the present paper. Absurd consequences of the standard solution 
          ---times of events in the travelling frame changing when a different arbitary
            choice of apatial coordinate system is made in the base frame, and abrupt (positive
             or negative) age changes of an object during rapid acceleration or deceleration--- 
            are pointed out.
           \par In Section 7 the DAE in the twin paradox experiment is discussed in 
             relation the Minkowski space-time plot. In this connection Minkowski's original 
             sign error when drawing the direction of the axes of space and time coordinates
             in the travelling frame S', pointed out in a recent paper by the present author
            ~\cite{JHFMinkP}, is important. This error remained uncorrected in all discussions
           of the twin paradox employing the Minkowski plot. Apart from this geometrical error,
            the fallacy in the standard  Minkowski plot analysis of the twin paradox is the 
           incorrect assumption that the TD effect of the physically independent reciprocal
            experiment is applicable in the primary experiment when the travelling twin observes
            the clock of her stay-at-home brother. Actually, as is obvious from inspection
            of the inverse LT for the primary experiment, the clock of the stay-at-home
            twin is observed to run faster (not slower, as in the reciprocal
            experiment) by the travelling twin. That this is the case is made perfectly
            clear by inspection of Eqns.(7.23)-(7.30).
            \par The twin paradox arises because of an apparent symmetry between configurations
              where the twin R is at rest and the twin T is in motion or the twin T is at rest and
            and the twin R is in motion. If only the relative motion of T and R is of physical
            significance the existence of such a symmetry appears to be well-founded. In the
              standard solution, the symmetry is broken by assuming the same relative velocity
              in the base and travelling frames but introducing `length contraction' of the
              spatial separation of the two twins (only while they are in relative motion) in
              the frame of the travelling twin. In the correct solution the spatial separation
               between the twins is the same at all corresponding instants in the base and
              travelling frames, but their relative velocity is different in the
              two frames. The physical basis of the symmetry breaking is the TD effect
              itself ---a moving clock is seen to run slow in comparison with an identical
              clock at rest in the observer's frame. The erroneous nature of the standard
             solution, in which R's clock is calculated to run slow when observed by T, results
               from the false identification of a configuration from the physically
                independent reciprocal experiment with an inverse transformation of the
                 primary experiment. 
                 \par That the primary and reciprocal experiments {\it are} physically
                independent has been previously pointed out in the literature. Builder~\cite{Builder}
                  made the essential point that the LT in the primary experiment where
                    $x' = 0$ and $x = vt$ is not the same (i.e. the space and time coordinate symbols have
                  a different physical meaning) as the LT of the reciprocal experiment where
                   $x = 0$ and $x' = -vt'$. Builder also correctly stated that identical results are
                   obtained by using (in either the primary or reciprocal experiment) the LT
             or its inverse  ---that is, in the primary experiment R's clock will be observed 
              by T to run faster than her own, not slower, as in the standard solution.
               In spite of this, Builder subsequently erroneously confused~\cite{Builder,Builder1}
           the base frame configuation
               of the reciprocal experiment with the travelling frame configuration of
               the primary experiment as described above.
           \par Another author who pointed out clearly the independence of the primary and
                reciprocal experiments was Max Born. Without considering the operational
              meaning of the coordinate symbols in the LT, Dingle~\cite{Dingle} had noticed the 
             antinomy similar to that of Eqns(4.16) and (4.17) above and concluded that the
              whole of special relativity theory is excluded by {\it reductio ad absurdum}.
              In his response~\cite{Born} to Dingle's direct request for comments, Born made
             the same correct observation as Builder ---in one case $x' = 0$ and $x = vt$, in the
             other $x = 0$ and $x' = -vt'$ so the symbols in the LT refer in the two cases
           to physically distinct experiments. 
              \par To the present writer's best knowledge, only one author~\cite{Swann} stated clearly
               in the published literature that the standard application
              of the generic LT (as the entries of Tables 1 and 2 and shown on the clocks with round
               dials in Figs. 9 and 11 above) does not correctly give the age of the
                travelling twin, as calculated by the TD relation, and shown on the clocks 
               with square dials in Figs. 9 and 11. A clear distinction was drawn between the
              age of the travelling twins and the times indicated by `synchronised' clocks
              but it was not pointed out that the latter are devoid of any physical meaning
            ---indeed, as shown in Section 6 above, correspond to patently absurd predictions.
               Qualitatively similar remarks to those in Ref.~\cite{Swann} are found at~\cite{TVF}.
             \par Many authors state, incorrectly, that the source if the asymmetry between the
              stay-at-home and travelling twins is that the travelling twin is accelerated 
              and decelerated, and therefore changes inertial frames, whereas the other stays
              always in the same inertial frame. In the space-billiard experiment described in Section 5
             above, the objects equivalent to R and T undergo identical acceleration and
             decleration programs, but the travelling object still ages less than the
             stay-at-home one. Langevin~\cite{Langevin} stated, correctly, that an inertial path
             linking two time-like separated space-time points corresponds to a maximum
              of the proper time integral between the two points, but incorrectly that this
             fact explains the DAE in the twin paradox experiment. Indeed in the 
             `three brothers' version of the twin paradox due to Lord Halsbury~\cite{Halsbury},
            described by Marder~\cite{MarderLHE}, where the clock setting of the brother 
            moving away from the Earth is transfered at the distant star to his brother
            moving towards the Earth at the same speed, the same global DAE effect occurs
            and no acceleration of deceleration occurs in the problem. 
            \par Other author's~\cite{Moller,FB,Perrin,JW,Gron,Styer} following Tolman~\cite{Tolman}
             have used the Equivalence Principle of general relativity to replace a constant proper
              frame acceleration of the travelling twin by an artificial `gravitational field' in
              the instantaneous rest frame of this twin. However all these authors made the same
              error of confusing the base frame of the reciprocal experiment with the travelling
              frame of the primary experiment, explained above in Section 7, when considering the
             phases of uniform motion of the round trip.
            \par Applying Occam's razor ---retaining only what is essential--- the present paper
             has not addressed the case of non-impulsive accelerations and decelerations, or the interesting
              analysis initiated by Langevin~\cite{Langevin} of exchange of light signals between
              the the twins during the voyage. Since, in the explanation of the DAE in the twin
              paradox given above, the acceleration and decleration phases of the motion
              play no role, the first restriction is of no importance for understanding the
              resolution of the twin paradox. The second subject is addressed by 
              another paper~\cite{JHFLangevin} containing a modern re-assesment of
              Langevin's original thought experiment~\cite{Langevin}.



\pagebreak

\end{document}